\documentclass[preprint]{aastex}
\usepackage{graphics,graphicx}
\usepackage{rotating}
\usepackage{amsmath}
\usepackage{txfonts}
\usepackage{natbib}

\shorttitle{Supernova PTF\,10bzf}

\shortauthors{A. Corsi et al.}

\begin{document}

\title{PTF\,10bzf (SN~2010ah): a broad-line Ic supernova discovered by the Palomar Transient Factory}
\author{A.~Corsi\altaffilmark{1}, E.~O.~Ofek\altaffilmark{2,3}, D.~A.~Frail\altaffilmark{4}, D.~Poznanski\altaffilmark{3,5,6}, I. Arcavi\altaffilmark{7} , A.~Gal-Yam\altaffilmark{7}, S.~R.~Kulkarni\altaffilmark{2}, K.~Hurley\altaffilmark{8}, P.~A.~Mazzali\altaffilmark{9,10}, D.~A.~Howell\altaffilmark{11,12}, M.~M.~Kasliwal\altaffilmark{2}, Y.~Green\altaffilmark{7}, D.~Murray\altaffilmark{11,12}, M.~Sullivan\altaffilmark{13}, D.~Xu\altaffilmark{7}, S.~Ben-ami\altaffilmark{7}, J.~S.~Bloom\altaffilmark{6}, S.~B.~Cenko\altaffilmark{6}, N.~M. Law\altaffilmark{14}, P.~Nugent\altaffilmark{5,6}, R.~M.~Quimby\altaffilmark{2}, V.~Pal'shin\altaffilmark{15}, J.~Cummings\altaffilmark{16}, V.~Connaughton\altaffilmark{17}, K.~Yamaoka\altaffilmark{18}, A.~Rau\altaffilmark{19}, W.~Boynton\altaffilmark{20}, I.~Mitrofanov\altaffilmark{21}, J.~Goldsten\altaffilmark{22}}
\altaffiltext{1}{LIGO laboratory, California Institute of Technology, MS 100-36, Pasadena, CA 91125, USA; \email{corsi@caltech.edu}}
\altaffiltext{2}{Cahill Center for Astrophysics, California Institute of Technology, Pasadena, CA, 91125, USA}
\altaffiltext{3}{Einstein Fellow}
\altaffiltext{4}{National Radio Astronomy Observatory, P.O. Box 0, Socorro, NM 87801, USA}
\altaffiltext{5}{Computational Cosmology Center, Lawrence Berkeley National Laboratory, 1 Cyclotron Road, Berkeley, CA 94720, USA}
\altaffiltext{6}{Department of Astronomy, 601 Campbell Hall, University of California, Berkeley, CA 94720-3411, USA}
\altaffiltext{7}{Department of Particle Physics and Astrophysics, The Weizmann Institute of Science, Rehovot 76100, Israel}
\altaffiltext{8}{Space Sciences Laboratory, University of California Berkeley, 7 Gauss Way, Berkeley, CA 94720, USA}
\altaffiltext{9}{INAF - Osservatorio Astronomico, vicolo dell'Osservatorio, 5, I-35122 Padova, Italy}
\altaffiltext{10}{Max-Planck Institut f\"{u}r Astrophysik, Karl-Schwarzschildstr. 1, D-85748 Garching, Germany}
\altaffiltext{11}{Las Cumbres Observatory Global Telescope Network, Inc, Santa Barbara, CA, 93117, USA}
\altaffiltext{12}{Department of Physics, University of California Santa Barbara, Santa Barbara, CA 93106, USA}
\altaffiltext{13}{Department of Physics (Astrophysics), University of Oxford, DWB, Keble Road, Oxford, OX1 3RH, UK}
\altaffiltext{14}{Dunlap Institute for Astronomy and Astrophysics, University of Toronto, 50 St. George Street, Toronto M5S 3H4, Ontario, Canada}
\altaffiltext{15}{Ioffe Physico-Technical Institute of the Russian Academy of Sciences, St. Petersburg, Russian Federation}
\altaffiltext{16}{University of Maryland, Baltimore County (UMBC), 1000 Hilltop Circle, Baltimore, MD 21250}
\altaffiltext{17}{University of Alabama in Huntsville CSPAR, Huntsville AL}
\altaffiltext{18}{Department of Physics and Mathematics, Aoyama Gakuin University, Kanagawa, Japan}
\altaffiltext{19}{Max-Planck-Institut f\"{u}r extraterrestrische Physik, Garching, Germany}
\altaffiltext{20}{University of Arizona, Department of Planetary Sciences, Tucson, AZ}
\altaffiltext{21}{Space Research Institute, Moscow, Russian Federation}
\altaffiltext{22}{Applied Physics Laboratory, Johns Hopkins University, Laurel, MD}

\begin{abstract}
We present the discovery and follow-up observations of a broad-line type-Ic supernova (SN), PTF\,10bzf (SN~2010ah), detected by the Palomar Transient Factory (PTF) on 2010 February 23. The SN distance is $\cong 218$\,Mpc, greater than GRB\,980425 / SN\,1998bw and GRB\,060218 / SN\,2006aj, but smaller than the other SNe firmly associated with gamma-ray bursts (GRBs). We conducted a multi-wavelength follow-up campaign with Palomar-48 inch, Palomar 60-inch, Gemini-N, Keck, Wise, \textit{Swift}, the Allen Telescope Array, CARMA, WSRT, and EVLA. Here we compare the properties of PTF\,10bzf with those of SN\,1998bw and other broad-line SNe. The optical luminosity and spectral properties of PTF~10bzf suggest that this SN is intermediate, in kinetic energy and amount of $^{56}$Ni, between non GRB-associated SNe like 2002ap or 1997ef, and GRB-associated SNe like 1998bw. No X-ray or radio counterpart to PTF\,10bzf was detected. X-ray upper-limits allow us to exclude the presence of an underlying X-ray afterglow as luminous as that of other SN-associated GRBs like GRB\,030329 or GRB\,031203. Early-time radio upper-limits do not show evidence for mildly-relativistic ejecta. Late-time radio upper-limits rule out the presence of an underlying off-axis GRB, with energy and wind density similar to the SN-associated GRB\,030329 and GRB\,031203. Finally, by performing a search for a GRB in the time window and at the position of PTF~10bzf, we find that no GRB in the IPN catalog could be associated with this SN.  
\end{abstract}


\section{Introduction}
\label{sec:Introduction}
The Palomar Transient Factory\footnote{http:$//$www.astro.caltech.edu$/$ptf$/$} \citep[PTF;][]{Law2009,Rau2009} is an on-going project optimized for detecting optical transients in the local Universe. One of its main objectives is the collection of a large sample of core-collapse supernovae \citep[SNe; e.g.][]{Arcavi2010}, for which multicolor optical light-curves and spectra can be obtained through dedicated follow-up resources. 

The explosive death of a SN massive progenitor occurs when its iron core collapses to a neutron star or a black hole. Core-collapse SNe are either of type Ib/Ic if the Hydrogen envelope of the progenitor is lost, or else of Type II \citep{Filippenko1997}. While the total kinetic energy released in the explosion is of the order of $10^{51}$\,erg, roughly the same as the energy of the jet that makes a gamma-ray burst (GRB), core-collapse SNe are in general not accompanied by highly relativistic mass ejection, and are visible from all angles.

The discovery of an association between a Ic SN and a long duration GRB in 1998 \citep{Galama1998,Kulkarni1998}, strongly supported the collapsar scenario \citep[e.g.][and references therein]{Mac1999,Meszaros2006,Woosley2006}. 
This was one of the most unusual Type Ic SNe seen up to that time: more luminous than typical Ic SNe, with broad lines and strong radio emission indicating a relativistic expansion speed \citep[$\Gamma \sim 2$; ][]{Kulkarni1998}. 

Since 1998, a total of five associations between GRBs and SNe have been spectroscopically confirmed (Table 1). GRBs with spectroscopically confirmed SNe are generally under-luminous and sub-energetic in comparison to typical long GRBs. A notable exception, though, is GRB\,030329 associated with SN\,2003dh, that represents the first solid evidence for a connection between ordinary GRBs and SNe \citep[][]{Garnavich2003,Hjorth2003,Kawabata2003,Matheson2003,Stanek2003}. 

After GRB\,030329, a SN-like brightening was reported at the position of GRB\,031203, and SN\,2003lw was then photometrically \citep{Bersier2004,Cobb2004,Malesani2004,GalYam2004}, and spectroscopically \citep{Malesani2004} confirmed to be associated with this GRB. Broad spectral features similar to SN\,1998bw also characterized SN\,2003lw. 

The understanding of the GRB-SN connection has further progressed thanks to discovery of GRB/XRF 060218 \citep{Pian2006,Modjaz2006,Sollerman2006,Ferrero2006}, associated with SN\,2006aj. The prompt spectrum of this GRB showed evidence for a typical non-thermal component (as observed also in ordinary GRBs), plus a thermal component. This last suggested that the shock breakout of an associated SN was for the first time being observed \citep{Campana2006,Waxman2007}.  A different central engine (a magnetar rather than a black hole), was also suggested by several authors to explain the long-duration and lower luminosity of GRB\,060218 \citep{Mazzali2006,Soderberg2006,Toma2007}. 

The breakout of a shock through the stellar surface is predicted to be the first electromagnetic signal marking the birth of a SN. The typical frequency of the emission is in the soft $\gamma$-rays for core-collapse SNe \citep[][]{Grassberg1971,Chevalier1976,Chevalier1992,Waxman2007,Chevalier2008,Katz2010,Nakar2010,Balberg2011,Katz2011,Nagakura2011,Rabinak2011}. Since XRF\,060218, several shock breakout candidate events have been proposed \citep{Soderberg2008,Gerzari2008,Schawinski2008,Modjaz2009,Ofek2010}. Multi-wavelength observations of core-collapse SN are fundamental to probe the breakout phase.

The most recent spectroscopically confirmed association between a GRB and a SN is represented by the case of GRB\,100316D \citep{Starling2010} and SN\,2010bh \citep{Bufano2010,Chornock2010,Wiersema2010,Cano2011}. GRB\,100316D was a long-duration, soft-spectrum GRB, resembling GRB\,060218 in these properties.

While the above mentioned observations have clearly established a connection between long-duration GRBs and type Ic core-collapse SNe, however, what makes some broad-line Ic SNe have an accompanying GRB is still a mystery. Some long GRBs are clearly not associated with a SN, e.g. GRB\,060505 \citep{Ofek2007} and GRB\,060614 \citep{DellaValle2006,Fynbo2006,Gal-Yam2006}. On the other hand, SN~1997ef remains so far the most energetic peculiar Ic SN without a clear GRB association \citep[e.g.][]{Mazzali2000}, and the broad-line Ic SN\,2009bb \citep{Soderberg2010} showed clear evidence of very high expansion velocity (as normally seen in GRB-related SNe), but no clear GRB association. However, it is remarkable that all GRB-related SNe discovered to date are broad line Ic SNe.

Theoretically, in order to produce a relativistic jet from a collapsing star to power the observed GRB, the stellar core has to carry a high angular momentum \citep{Woosley2006a}, so that the spin axis provides a natural preferred propagation direction for the jet. Moreover, the jet needs to be ``clean'', with small baryon contamination, so that it can achieve a relativistic speed, with Lorentz factor $\Gamma$ typically greater than 100 \citep{Piran1999,Lithwick2001,Liang2010,Abdo2009a,Abdo2009b,Abdo2009c,Corsi2010}. The outflow also needs to be collimated, with an aperture angle of the order of $1^{\circ} - 10^{\circ}$ for bright GRBs \citep{Frail2001,Bloom2003,Liang2008,Racusin2009}. 

Several types of GRB central engine have been discussed, the leading candidate being a black hole plus torus system \citep{Woosley1993,MacFadyen1999,Proga2003,Zhang2003}. An alternative candidate is a rapidly spinning, highly magnetized neutron star \citep[magnetar, e.g.][]{Usov1992,Thompson2004,Bucciantini2008,Metzger2011}. In the black hole scenario, the first energy source is the accretion power from the torus. On the other hand, the main power of a millisecond magnetar engine is its spin down power.

Whatever the nature of the GRB central engine is, it has to generate both a narrowly collimated, highly relativistic jet to make the GRB, and a wide angle, sub-relativistic outflow responsible for exploding the star and making the SN. To some extent, the two components may vary independently, so it is possible to produce a variety of jet energies and SN luminosities. \citet{Woosley2007} have shown that at least a $\sim10^{48}$ erg s$^{-1}$ power is required for a jet to escape a massive star before that star either explodes or is accreted, and lower energy and ``suffocated'' bursts may be particularly prevalent when the metallicity is high, i.e., in the modern universe at low redshift.

A key prediction of jet models for GRBs \citep[e.g.][]{Piran2004}, combined with the association of long GRBs with type Ib/Ic core-collapse SNe \citep[e.g.][]{Woosley2006}, is that some (spherical) SN explosions will be accompanied by off-axis GRBs, whose gamma-ray signal is missed because the jet is not pointed at us, but whose afterglow emission could be visible at lower energies (from radio to X-rays), once the jet decelerates. These are also referred to as ``orphan afterglows''. Detecting an off-axis GRB would constitute a direct proof of the popular jet model for GRBs but, after a decade of efforts \citep{Berger2003,GalYam2006,Soderberg2006}, such a detection has not yet been achieved. Orphan afterglows may also be produced by ``dirty fireballs'', i.e., cosmological fireballs whose ejecta carry too many baryons to produce a GRB. Multi-color, wide-area searches, combined with radio follow-up, can help distinguishing between off-axis GRBs and dirty fireballs \citep{Perna1998,Rhoads2003}.

Based on radio follow-up campaigns of Ib/c SNe \citep{Berger2003,Soderberg2010}, it is now believed that no more than 1\% of SNe type Ib/c are powered by central engines. These observational facts are giving us hints at the fundamental role that broad-line Ic SNe play in engine-drive explosions. GRBs are known to prefer low-metallicity dwarf galaxies \citep[e.g.][]{Fynbo2003,Fruchter2006}, and to be associated only with a particular subclass of core-collapse SNe: broad-line Ic events. Previous searches, however, targeted mostly regular SNe Ib/Ic, residing in giant, high-metallicity hosts. This is why broad-line Ic events deserve to be studied with particular attention \citep{Modjaz2008}. 

Multi-wavelength follow-up campaigns of broad-line Ib/c SNe are especially important to probe high expansion velocities, and/or any associated GRB. Moreover, because of their low (electromagnetic) luminosity compared to GRBs, usually SNe are observed at non-cosmological distances ($z\lesssim1$). Thus, broad-line Ib/c also represent a tool to search for nearby GRB explosions and understand their connection with cosmological GRBs \citep[e.g.][]{Norris2002,Pod2004,Guetta2007,Liang2007,Virgili2009}. 

In this paper, we present the discovery of a broad-line Ic SN, PTF\,10bzf (SN\,2010ah), detected by PTF. This event is interesting because of two reasons. First, as a broad-line SN located at a distance smaller than most GRB-associated events (except for GRB\,980425 / SN\,1998bw and GRB\,060218 / SN\,2006aj, see Table 1). Next, this event enjoyed a rich radio-to-X-ray follow-up campaign. In what follows, we first describe the observations that led to the discovery of PTF\,10bzf (\S 2), and its multi-wavelength follow-up campaign (\S 3). Then, we compare PTF\,10bzf with SN\,1998bw and other GRB-associated SNe, and describe the results of an associated GRB search (\S 4). Finally, we conclude in \S 5. 

\begin{deluxetable}{lllccc}
\tablecolumns{6}
\tablewidth{0pc}
\tablecaption{Summary of light-curve properties}
\tablehead{
\colhead{SN} &
\colhead{type} &
\colhead{Associated} &
\colhead{$d_{\rm L}$} &
\colhead{$M_{\rm R}$} &
\colhead{$M_{Ni}/M_{\odot}$}\\
\colhead{} &
\colhead{} &
\colhead{GRB} &
\colhead{Mpc}&
\colhead{(mag)} &
\colhead{}
}
\startdata
SN\,1998bw\tablenotemark{a} & engine-driven BL-Ic & GRB\,980425 & 37& $-19.36\pm0.05$ & $0.4-0.5$\\
SN\,2003dh\tablenotemark{b} & engine-driven BL-Ic & GRB\,030329 & 810& $\approx -19$&  $0.25-0.45$ \\
SN\,2003lw\tablenotemark{c} & engine-driven BL-Ic & GRB\,031203 & 477&  $-19.90\pm0.08$ &  $0.45-0.65$\\
SN\,2006aj\tablenotemark{d} & engine-driven BL-Ic & GRB\,060218 & 140& $-18.81\pm0.06$ & $0.21$  \\
SN\,2010bh\tablenotemark{e} & engine-driven BL-Ic & GRB\,100316D & $261$& - & - \\
SN\,2009bb\tablenotemark{f} & engine-driven BL-Ic & none & 40 & $-18.56\pm0.28$ & 0.16-0.28\\
\hline
SN\,2003jd\tablenotemark{g} & BL-Ic & none & 78& $-18.94\pm0.30$ & 0.26-0.45\\
SN\,2002ap\tablenotemark{h} & BL-Ic & none & 7.8& $-17.50\pm0.32$ & 0.06-0.12
\enddata
\tablecomments{See also \citet{Woosley2006}, and references therein, for a recent review.}
\tablenotetext{a}{\citet{Galama1998,Iwamoto1998,Nakamura2001}}
\tablenotetext{b}{\citet{Hjorth2003,Matheson2003,Deng2005}}
\tablenotetext{c}{\citet{Malesani2004,Mazzali2006}}
\tablenotetext{d}{\citet{Mazzali2006a,Pian2006,Soderberg2006,Valenti2008}}
\tablenotetext{e}{\citet{Starling2010}}
\tablenotetext{f}{\citet{Pignata2010}}
\tablenotetext{g}{\citet{Mazzali2005,Valenti2008,Drout2010}}
\tablenotetext{h}{\citet{GalYam2002,Mazzali2002,Foley2003,Drout2010}}
\end{deluxetable}

\section{Observations and data reduction}
\label{sec:Observations}
\begin{figure}
\begin{center}
\includegraphics[width=8.5cm]{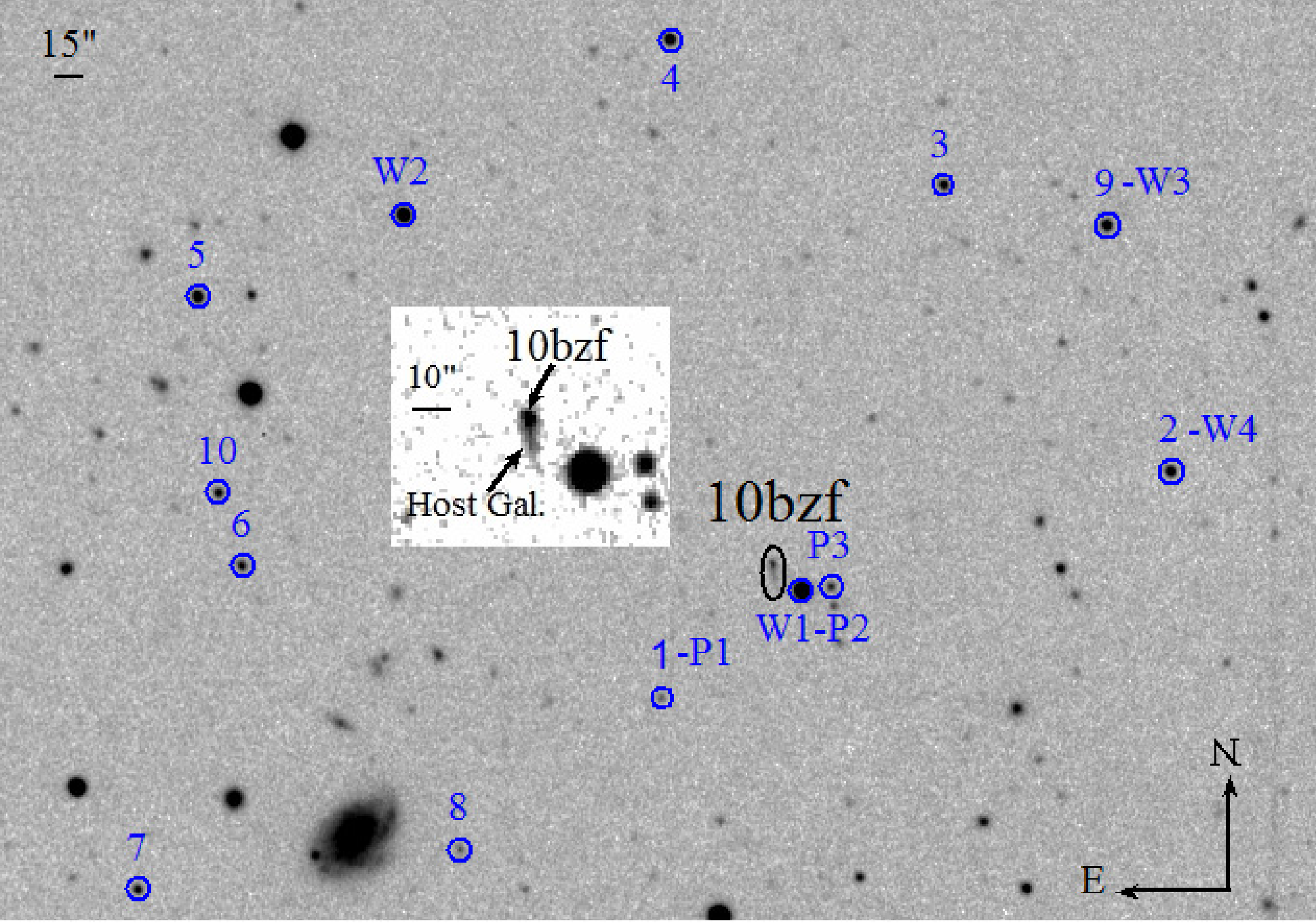}
\caption{P48 discovery image of PTF\,10bzf in the $R$-band. For clarity purposes, a circle of $5''$ radius marks the position of the ten reference stars used for P48 photometry calibration (1-10), and of the four reference stars used for calibration of the Wise data (W1-W4). The stars labeled as P1-P3 were used for the calibration of P60 photometry. The position of PTF\,10bzf and its host galaxy are enclosed by a black ellipse. The inset shows in more detail PTF\,10bzf inside its host. The SN is located at RA=11:44:02.99 and Dec=+55:41:27.6 (J2000). \label{zoom}}
\end{center}
\end{figure}

On 2010 February 23.5038 (hereafter all times are given in UTC), we discovered a broad-line type-Ic SN, PTF\,10bzf, visible at a magnitude of $R\approx18.86$ (see Table 2 and Figure \ref{zoom}), in a $60$\,s exposure image taken with the Palomar 48-inch telescope (P48). The SN was not seen in previous images of the same field taken on 2010 February 19.4392, down to a limiting magnitude of $R > 21.3$. The SN J2000 position is RA=11:44:02.99, Dec=+55:41:27.6 \citep{ATEL2470}, $\cong 5''.2$ offset, and at a position angle of $\cong 5$\,deg (North through East) about the position of the galaxy SDSS\footnote{Sloan Digital Sky Survey \citep{York2000}.} J114402.98+554122.5. 

Digital copies of our data can be downloaded directly from the Weizmann Institute of Science Experimental Astrophysics Spectroscopy System \citep[WISEASS\footnote{http:$//$www.weizmann.ac.il$/$astrophysics$/$wiseass$/$},][]{Yaron2011}.

\subsection{Optical photometry}
\begin{figure*}
\begin{center}
\includegraphics[height=8.5cm]{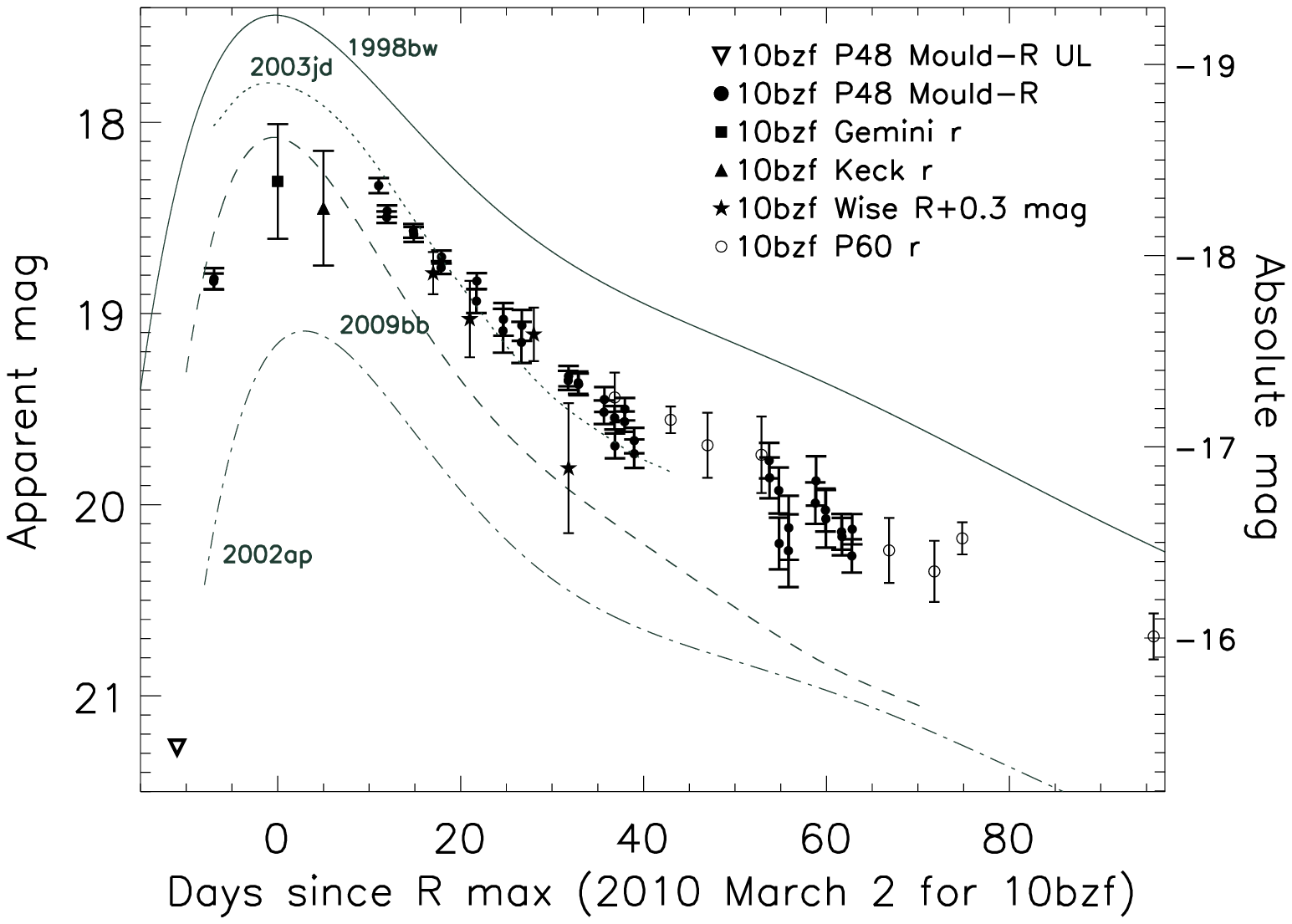}
\caption{Light-curve of PTF\,10bzf corrected for Galactic extinction. Absolute magnitudes are plotted on the right axis. P48 data in Mould-$R$ filter (filled circles), and P60 data in $r$-band (open circles) are calibrated using $r$-band magnitudes of SDSS stars, and expressed in the AB system. For P48, we also include color term corrections estimated considering the $r-i$ SDSS colors of the reference stars. Synthetic photometry obtained from Gemini (filled square) and Keck (filled triangle) spectra is referred to SDSS $r$-band and expressed in AB system. From the Gemini spectrum, we estimate $R-r\approx-0.1$\,mag for the conversion from SDSS $r$-band magnitudes in AB system, to $R$-band magnitudes in Vega system. Wise data in the $R$ filter and Vega system (stars), are calibrated to SDSS as described in \citet{Jordi2006}. For comparison with P48 photometry, Wise data are shifted to account for $r-R\approx 0.3$\,mag for the conversion from SDSS $r$-band magnitudes in AB system, to Wise $R$-band magnitudes in Vega system \citep{Jordi2006}. We also plot the $R$-band light-curve template of SN\,1998bw (solid line), SN\,2003jd (dotted line), SN\,2009bb (dashed line), and SN\,2002ap (dash-dotted line), rescaled to the redshift of PTF\,10bzf. These templates are obtained interpolating data retrieved from: \citet{Galama1998} for SN\,1998bw; \citet{GalYam2002,Foley2003} for SN\,2002ap; \citet{Valenti2008} for SN\,2003jd; \citet{Pignata2010} for SN\,2009bb.
\label{Fig2}}
\end{center}
\end{figure*}

P48 observations of the PTF\,10bzf field were performed with the Mould-$R$ filter. A high-quality image produced by stacking several images of the same field (obtained between May 2009 and June 2009), was used as a reference and subtracted from the individual images. Photometry was performed relative to the $r$-band magnitudes of ten SDSS reference stars in the field, including $r$-$i$ color term corrections, using an aperture of 2 arc sec radius, and applying aperture corrections to account for systematic errors and errors introduced by the subtraction process. Aperture corrections are all below $\approx 0.04$ mag. All the P48 photometry is listed in Table \ref{Tab1}. The P48 calibrated light-curve of PTF\,10bzf is plotted in Figure \ref{Fig2}.

P60 observations were carried out in the $B$, $g$, $r$, $i$, $z$ bands. We also observed the PTF\,10bzf field with the 1\,m telescope at the Wise observatory\footnote{http:$//$wise$-$obs.tau.ac.il$/$} using $BVRI$ filters (see Table 2). For both P60 and Wise observations, image subtraction was performed using the common point spread function method via the ``mkdifflc'' routine \citep{GalYam2004,GalYam2008}. Errors on the P60 and Wise data are estimated by using ``artificial'' sources at a brightness similar to that of the real SN, with the scatter in their magnitudes providing an estimate of the error due to subtraction residuals. The measured magnitude was calibrated against the magnitudes of SDSS stars in the same field (see Figure 1). The calibration procedure described in \citet{Jordi2006} was used for the $BVRI$ data. Calibration errors were summed in quadrature with the subtraction errors.

\begin{deluxetable}{lllccr}
\tablecolumns{6}
\tablewidth{0pc}
\tablecaption{PTF\,10bzf follow-up campaign.\label{Tab1}}
\tablehead{
\colhead{JD-2455251.004} &
\colhead{Telescope} &
\colhead{$\Delta t$} &
\colhead{Band} &
\colhead{Mag or Flux} &
\colhead{Reference}  \\
\colhead{(days since Feb 23.504)}&
\colhead{} &
\colhead{(s)} &
\colhead{} &
\colhead{} &
\colhead{} 
}
\startdata
 -4.061 & P48 & 60 & Mould-$R$  & $>21.3$  & ATEL 2470\\ 
 $0.000$& P48 & 60 &  Mould-$R$ & $18.863\pm0.041$  
& ATEL 2470\\
 0.045& P48 & 60 &  Mould-$R$ & $18.850\pm0.056$  
& This paper\\
 18.038& P48 &60 & Mould-$R$  & $18.362\pm0.040$  
& This paper \\
 18.921& P48 & 60 & Mould-$R$ & $18.527\pm0.030$  
& This paper\\
 18.966& P48 & 60 & Mould-$R$ & $18.495\pm0.030$  
&This paper\\
 21.832& P48 & 60 & Mould-$R$ & $18.599\pm0.036$  
& This paper\\
 21.876& P48 & 60 & Mould-$R$ & $18.618\pm0.039$  
& This paper\\
 24.866& P48 & 60 & Mould-$R$ & $18.791\pm0.033$ 
& This paper\\
 24.911& P48 & 60 & Mould-$R$ & $18.735\pm0.034$ 
& This paper\\
 28.727& P48 & 60 & Mould-$R$ & $18.966\pm0.062$ 
&This paper \\
 28.771& P48 & 60 & Mould-$R$ & $18.861\pm0.042$ 
&This paper\\
 31.636& P48 & 60 & Mould-$R$ & $19.12\pm0.11$  
&This paper\\
 31.679& P48 & 60 & Mould-$R$ & $19.061\pm0.085$ 
&This paper\\
 33.650& P48 & 60 & Mould-$R$ & $19.18\pm0.11$ 
&This paper\\ 
 33.693& P48 & 60 & Mould-$R$ & $19.093\pm0.081$ 
&This paper\\
 38.758& P48 & 60 & Mould-$R$ & $19.381\pm0.050$ 
&This paper\\
 38.802& P48 & 60 & Mould-$R$ & $19.359\pm0.053$ 
&This paper\\
 39.857& P48 & 60 & Mould-$R$ & $19.393\pm0.058$ 
&This paper\\
 39.901& P48 & 60 & Mould-$R$ & $19.401\pm0.057$ 
&This paper\\
 42.670& P48 & 60 & Mould-$R$ & $19.547\pm0.062$ 
&This paper\\
 42.713& P48 & 60 & Mould-$R$ & $19.480\pm0.065$ 
&This paper\\
 43.829& P48 & 60 & Mould-$R$ & $19.576\pm0.061$ 
&This paper\\
 43.873& P48 & 60 & Mould-$R$ & $19.723\pm0.065$ 
&This paper\\
 44.926& P48 & 60 & Mould-$R$ & $19.597\pm0.053$ 
&This paper\\
 44.962& P48 & 60 & Mould-$R$ & $19.531\pm0.059$ 
&This paper\\
 45.983& P48 & 60 & Mould-$R$ & $19.696\pm0.067$ 
&This paper\\
 45.984& P48 & 60 & Mould-$R$ & $19.764\pm0.075$ 
&This paper\\
 60.737& P48 & 60 & Mould-$R$ & $19.800\pm0.093$ 
&This paper\\ 
 60.781& P48 & 60 & Mould-$R$ & $19.89\pm0.11$ 
&This paper\\
 61.783& P48 & 60 & Mould-$R$ & $19.96\pm0.12$ 
&This paper\\
 61.826& P48 & 60 & Mould-$R$ & $20.23\pm0.13$ 
&This paper\\
 62.835& P48 & 60 &Mould-$R$ & $20.27\pm0.19$ 
&This paper\\
 62.878& P48 & 60 & Mould-$R$ & $20.15\pm0.17$ 
&This paper\\
 65.782& P48 & 60 & Mould-$R$ & $20.02\pm0.11$ &This paper\\
 65.857& P48 & 60 & Mould-$R$ & $19.91\pm0.13$ &This paper\\
 66.888& P48 & 60 & Mould-$R$ & $20.06\pm0.11$ &This paper\\
 66.932& P48 & 60 & Mould-$R$ & $20.10\pm0.15$ &This paper\\
 68.649& P48 & 60 & Mould-$R$ & $20.173\pm0.093$ &This paper\\
 68.692& P48 & 60 & Mould-$R$ & $20.198\pm0.098$  &This paper\\
 69.771& P48 & 60 & Mould-$R$ & $20.299\pm0.087$ &This paper\\
 69.815& P48 & 60 & Mould-$R$ & $20.159\pm0.079$ &This paper\\
\hline
 43.840 & P60 &180 & $r$ & $19.47\pm0.13$ & This paper\\
 43.841 & P60 &180 & $i$ & $19.51\pm0.32$ & This paper \\
 44.898 & P60 &180 & $z$ & $19.86\pm0.57$ & This paper\\
 44.901 & P60 &180 & $B$ & $21.63\pm0.75$\tablenotemark{a} & This paper\\
 49.960 & P60 &180 & $g$ & $21.17\pm0.17$ & This paper \\
 49.962 & P60 &180 & $r$ & $19.587\pm0.070$ & This paper\\
 49.964 & P60 &180 & $i$ & $19.746\pm0.044$ & This paper\\
 49.965 & P60 &180 & $z$ & $18.99\pm0.31$ & This paper\\
 49.967 & P60 &180 & $B$ & $21.7\pm1.0$\tablenotemark{a} & This paper\\
 53.993 & P60 &180 & $r$ & $19.72\pm0.17$ & This paper\\
 59.895 & P60 &180 & $r$ & $19.77\pm0.20$ & This paper\\
 65.904 & P60 &180 & $i$ & $19.96\pm0.17$ & This paper\\
 73.842 & P60 &180 & $i$ & $20.15\pm0.17$ & This paper\\
 73.844 & P60 &180 & $r$ & $20.27\pm0.17$ & This paper\\
 78.790 & P60 &180 & $r$ & $20.38\pm0.16$ & This paper\\
 79.790 & P60 &180 & $i$ & $20.50\pm0.17$ & This paper\\
 79.794 & P60 &180 & $B$ & $21.82\pm0.83$\tablenotemark{a} & This paper\\
 79.797 & P60 &180 & $g$ & $20.88\pm0.16$ & This paper\\
 80.771 & P60 &180 & $z$ & $19.47\pm0.45$ & This paper\\
 81.848 & P60 &180 & $r$ & $20.207\pm0.084$ & This paper\\
 99.798 & P60 &180 & $i$ & $20.99\pm0.36$ & This paper \\
 99.801 & P60 &180 & $B$ & $22.5\pm1.3$\tablenotemark{a} & This paper\\
 99.803 & P60 &180 & $g$ & $21.27\pm0.19$ & This paper\\
 102.774 & P60 &180 & $r$ & $20.72\pm0.12$ & This paper\\
 102.776 & P60 &180 & $B$ & $22.45\pm0.81$\tablenotemark{a} & This paper\\
 102.780 & P60 &180 & $g$ & $21.66\pm0.31$ & This paper \\
 103.791 & P60 &180 & $i$ & $20.80\pm0.27$ & This paper \\
\hline
 24 & Wise (PI) & 600 & $B$ & $20.34\pm0.57$\tablenotemark{b} & This paper\\
 24 & Wise (PI) & 600 & $V$ & $19.08\pm0.18$\tablenotemark{b} & This paper\\
 24 & Wise (PI) & 600 & $R$ & $18.55\pm0.11$\tablenotemark{b} &This paper \\
 28 & Wise (PI) & 600 & $I$ & $18.42\pm0.20$\tablenotemark{b} & This paper \\
 28 & Wise (PI) & 600 & $B$ & $20.60\pm0.64$\tablenotemark{b}& This paper\\
 28 & Wise (PI) & 600 & $V$ & $19.36\pm0.27$\tablenotemark{b} & This paper\\
 28 & Wise (PI) & 600 & $R$ & $18.79\pm0.20$\tablenotemark{b} & This paper \\
 28 & Wise (PI) & 600 & $I$ & $18.60\pm0.13$\tablenotemark{b} & This paper \\
 35 & Wise (LAIWO) & 720 & $V$ & $20.27\pm0.42$\tablenotemark{b} & This paper\\
 35 & Wise (LAIWO) & 720 & $R$ & $18.87\pm0.14$\tablenotemark{b} & This paper\\
 35 & Wise (LAIWO) & 720 & $I$ & $19.00\pm0.61$\tablenotemark{b} & This paper \\
 38 & Wise (LAIWO) & 720 & $V$ & $19.87\pm0.69$\tablenotemark{b} & This paper\\
 38 & Wise (LAIWO) & 720 & $R$ & $19.57\pm0.34$\tablenotemark{b} & This paper\\
\hline
 7 & Gemini & 450 & $g$\tablenotemark{c} & $18.6\pm0.3$ &   ATEL 2470\\
 7 & Gemini & 450 & $r$\tablenotemark{c} & $18.3\pm0.3$ &   ATEL 2470\\
 7 & Gemini & 450 & $i$\tablenotemark{c} & $18.7\pm0.3$ &   ATEL 2470\\
\hline
 12 & Keck  & 240&$g$\tablenotemark{c} & $18.91\pm0.3$ &  This paper\\
 12 & Keck  & $2\times80$&$r$\tablenotemark{c} & $18.48\pm0.3$ &  This paper\\
 12 & Keck  & $2\times80$&$i$\tablenotemark{c} & $18.70\pm0.3$ &  This paper\\ 
\hline
 8.52 & UVOT & $5\times10^{3}$ & $B$   & $18.73\pm0.10$ 
& ATEL 2471\\
 8.52 & UVOT & $5\times10^{3}$ & $U$   & $18.88\pm0.12$ & ATEL 2471 \\
 8.52 & UVOT & $5\times10^{3}$ & $UVW1$   & $20.07\pm0.18$ & ATEL 2471 \\
 8.52 & UVOT & $5\times10^{3}$ & $UVW2$   & $20.18\pm0.26$ & ATEL 2471\\
 12.80 & UVOT  & $2.5\times10^{3}$ & $U$   & $19.68\pm0.14$  & ATEL 2471\\
 12.80 & UVOT  & $2.5\times10^{3}$ & $UVW1$   & $20.12\pm0.24$ & ATEL 2471\\
\hline
8.52 & XRT  & $5\times10^{3}$ &$0.3-10$ keV   &  $< 1.3\times10^{-14}$ erg~s$^{-1}$cm$^{-2}$& ATEL 2471\\
12.80 & XRT  & $2.5\times10^{3}$ & $0.3-10$ keV  &$< 2.7\times10^{-14}$ erg~s$^{-1}$cm$^{-2}$& ATEL 2471\\
\hline
9.69 & CARMA & $19.8\times10^{3}$ & $95$ GHz & $(-3.7\pm1.8)\times10^{3}$ $\mu$Jy& ATEL 2473\\
\hline
 10.14 & Allen  & $4.8\times10^{3}$ & $3.09$ GHz & $<1.5\times10^{3}$ $\mu$Jy& ATEL 2472\\ 
\hline
 17.69 & EVLA &  $5.76\times10^{3}$& $4.96$ GHz &  $<33$ $\mu$Jy& ATEL 2483\\
 86.71 & EVLA & $3600$ & 6 GHz & $<36\mu$Jy& This paper\\
 276.9  & EVLA & $7200$ & 4.96 GHz & $<35\mu$Jy&This paper\\
\hline
18.24 &WSRT&  $28.8\times10^{3}$& $4.8$~GHz & $<126$ $\mu$Jy& ATEL 2479
\enddata
\tablenotetext{a}{$B$ magnitudes (in Vega system) calibrated to SDSS using the conversions described in \citet{Jordi2006}.}
\tablenotetext{b}{$BVRI$ magnitudes (in Vega system) calibrated to SDSS using the conversions described in \citet{Jordi2006}.}
\tablenotetext{c}{Synthetic photometry is referred to SDSS $g$, $r$, $i$ filters (in AB system). Errors are dominated by flux calibration errors.}
\tablecomments{Magnitudes are not corrected for Galactic extinction \citep[$E(B-V)=0.012$\,mag;][]{Schlegel1998}. P48 and P60 observations are calibrated to SDSS (which is estimated to be on the AB system within $\pm0.01$\,mag in the $r$, $i$ and $g$ bands; within $\pm0.02$\,mag in the $z$ band). P48 errors include (in quadrature) zeropoint calibration errors ($\lesssim 0.01$\,mag), color term errors ($\lesssim 0.03$\,mag), and aperture correction errors ($\lesssim 0.04$\,mag). For P48 detections, positive counts were collected in the subtracted images at the SN position and magnitudes are above the image $3\sigma$ limiting magnitude. All upper-limits are at $3\sigma$.}
\end{deluxetable}

\subsection{Spectroscopy}
\begin{figure}
\begin{center}
\includegraphics[width=8.5cm]{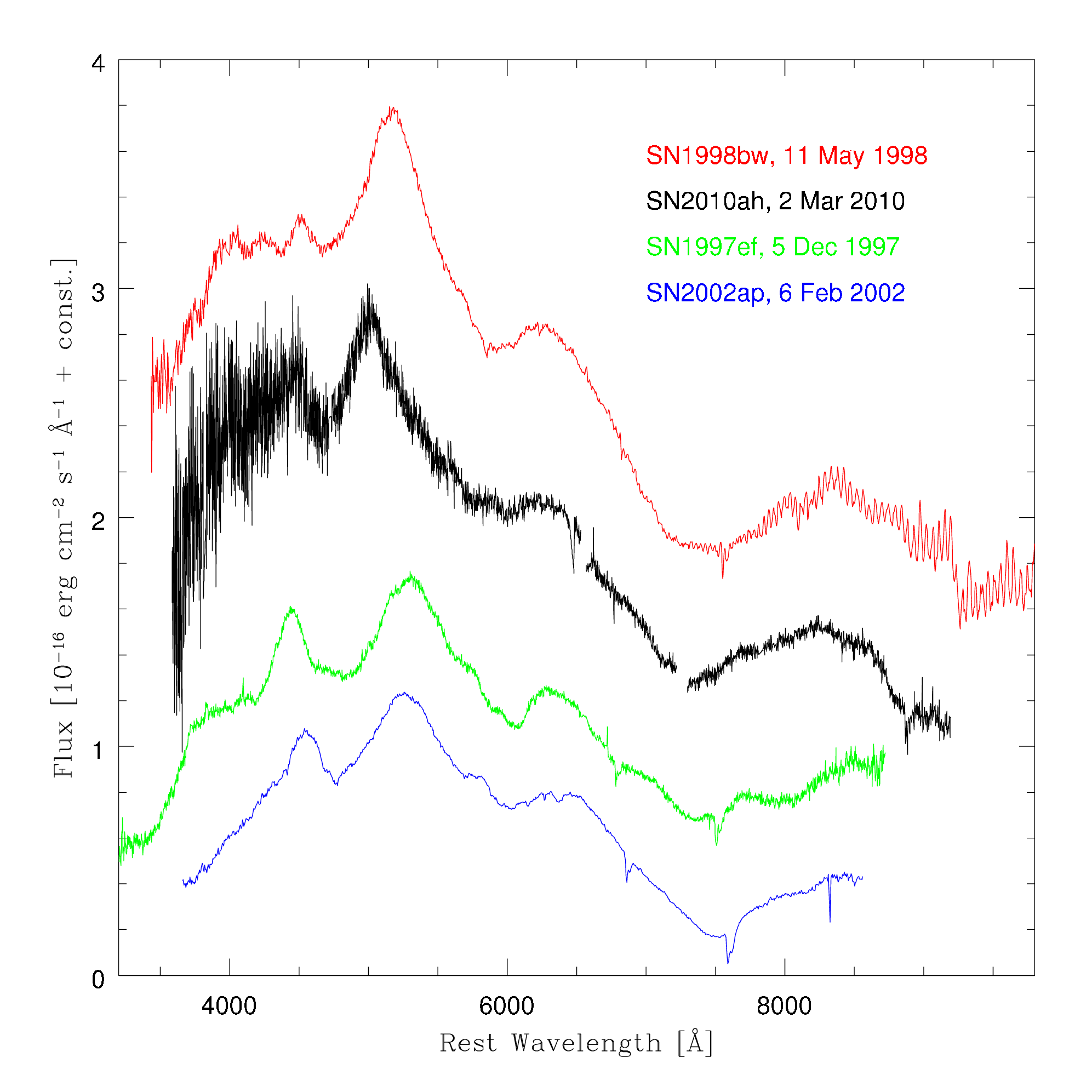}
\caption{Gemini (black) spectrum of PTF\,10bzf obtained on 2010 March 2, around the time of the $R$-band maximum. The spectrum is compared with that of SN~1997ef (green), SN~1998bw (red), SN~2002ap (blue) at similar epochs.  The SNe are ordered by peak luminosity. This ordering reveals a sequence also in line velocity and blending. The spectrum of SN\,1998bw is the most affected by line broadening, as seen for example in the almost complete absence of the re-emission peak near 4500 \AA\ and by the complete blending of the \ion{O}{1} 7773 and the \ion{Ca}{2} IR triplet. PTF\,10bzf is intermediate between SN\,1998bw and the two non-GRB SNe 1997ef and 2002ap in both of these respects. Spectral data were retrieved from: \citet{Mazzali2000} for SN~19997ef; \citet{Patat2001} for SN~1998bw; \citet{Mazzali2002} for SN~2002ap. Main telluric lines have been removed from PTF\,10bzf spectrum.\label{Gemini}}
\end{center}
\end{figure}

\begin{figure}
\begin{center}
\includegraphics[width=8.5cm]{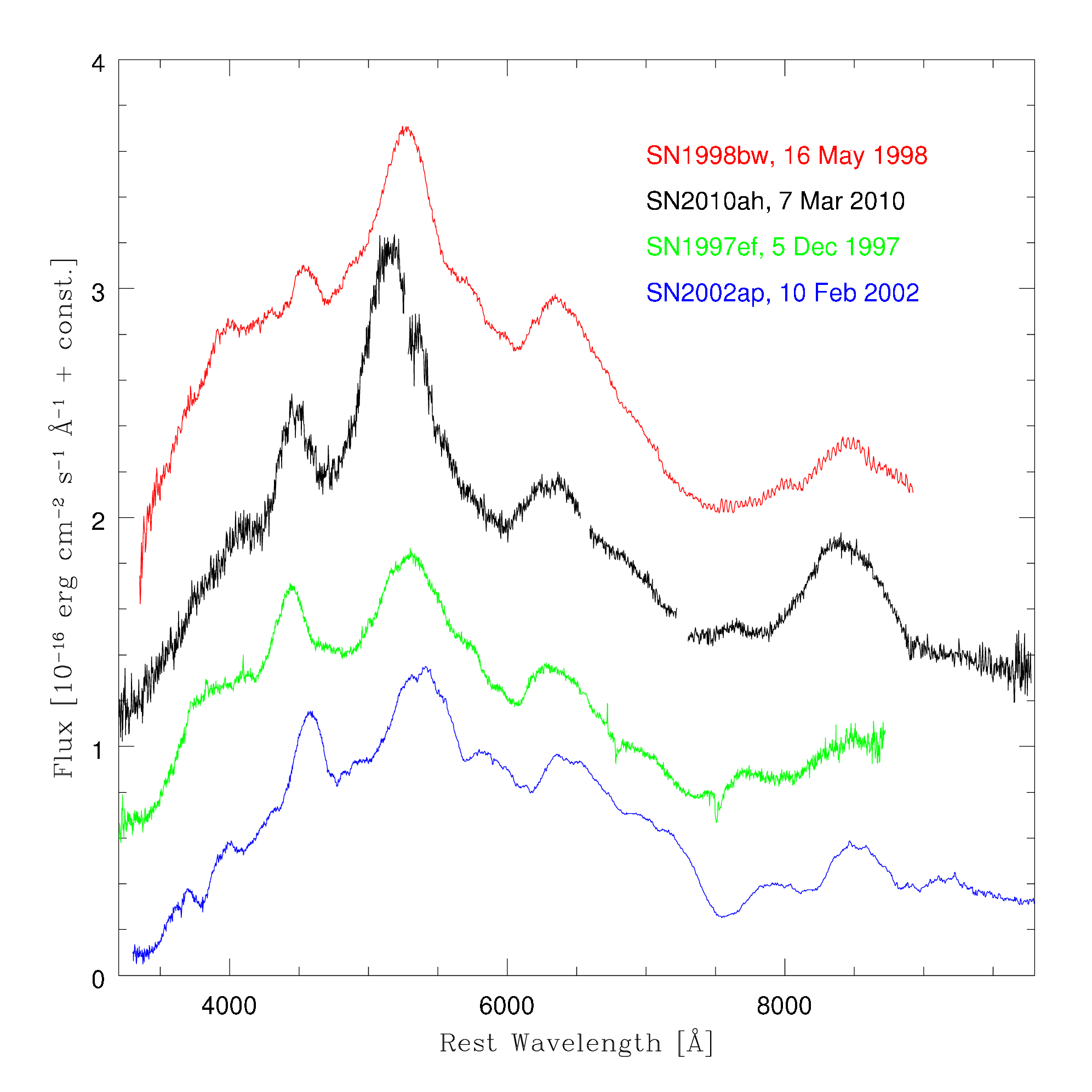}
\caption{Keck (black) spectrum of PTF\,10bzf obtained on 2010 March 7, about 5 days post $R$-band maximum. The spectrum is compared with that of SN~1997ef (green), SN~1998bw (red), SN~2002ap (blue) at similar epochs.  The SNe are ordered by peak luminosity. This ordering reveals a sequence also in line velocity and blending. The spectrum of SN\,1998bw is the most affected by line broadening, as seen for example in the almost complete absence of the re-emission peak near 4500 \AA\ and by the complete blending of the \ion{O}{1} 7773 and the \ion{Ca}{2} IR triplet. PTF\,10bzf is intermediate between SN\,1998bw and the two non-GRB SNe 1997ef and 2002ap in both of these respects. Spectral data were retrieved from: \citet{Mazzali2000} for SN~19997ef; \citet{Patat2001} for SN~1998bw; \citet{Mazzali2002} for SN~2002ap. Main telluric lines have been removed from PTF\,10bzf spectrum.\label{Keck}}
\end{center}
\end{figure}

Gemini North GMOS\footnote{The Gemini Observatory is operated by the Association of Universities for Research in Astronomy, Inc., under a cooperative agreement with the NSF on behalf of the Gemini partnership: the National Science Foundation (United States), the Science and Technology Facilities Council (United Kingdom), the National Research Council (Canada), CONICYT (Chile), the Australian Research Council (Australia), Minist\'{e}rio da Ci\^{e}ncia e Tecnologia (Brazil) and Ministerio de Ciencia, Tecnolog\'{i}a e Innovaci\'{o}n Productiva (Argentina).} \citep{Hook2004} spectra were taken on 2010 March 2 (Program ID: GN-2010A-Q-20), using a $1''$ slit, with the B600 and R400 gratings. The observations had an exposure time of 450 s (see Table 2), the airmass was 1.236, the sky position angle was $191^{\circ}$, and the standard star was Feige34. Standard data reduction was performed with IRAF V2.14, using the Gemini 1.10 reduction packages. The Gemini spectrum of PTF\,10bzf is shown in Figure \ref{Gemini}. A redshift of $z=0.0498\pm0.0003$ was derived from the host galaxy emission lines of \ion{O}{3}, H$\alpha$, H$\beta$, \ion{N}{2}, and \ion{S}{2}. The spectrum (black line in Figure \ref{Gemini}), resembles that of SN\,1998bw at a similar epoch \citep[red line in Figure \ref{Gemini} and][]{Galama1998}, and shows very broad lines, leading us to classify this SN as a broad-line Ic.

PTF\,10bzf was also observed by Keck I/LRIS\footnote{The W.M. Keck Observatory is operated as a scientific partnership among the California Institute of Technology, the University of California and the National Aeronautics and Space Administration. The Observatory was made possible by the generous financial support of the W.M. Keck Foundation.} using a $1''$ slit, with the 400/8500 grating plus 7847\,\AA\ central wavelength on the red side, and with the 400/3400 grism on the blue side. The exposure time was $2\times 80$\,s on the red side, and $1\times240$\,s on the blue side (see Table 2). Keck data were reduced using the standard longslit reduction packages developed in the IRAF environment (Figure \ref{Keck}).

Both Gemini and Keck spectra were used to derive synthetic photometry in SDSS $r$-, $g$-, and $i$-bands (Table 2). Synthetic photometry was performed as described in \citet{Poznanski2002}.

\subsection{\textit{Swift} follow-up observations}
A two-epoch observation of PTF\,10bzf was performed as part of a \textit{Swift} Target of Opportunity program\footnote{``Unveiling New Classes of Transients with Palomar Transient Factory'', PI S. R. Kulkarni}. \textit{Swift}/XRT did not detect any X-ray counterpart to PTF\,10bzf \citep{ATEL2471}. The corresponding upper limits are reported in Table 2, where we have converted the 0.3--10\,keV XRT count rates into fluxes assuming a photon index of $\Gamma=2$, and correcting for Galactic absorption (N$_{\rm H}\approx10^{20}$\,cm$^{-2}$).

During the first epoch, PTF\,10bzf was detected by \textit{Swift}/UVOT \citep{ATEL2471}, and was observed to fade in the subsequent observation. The measured magnitudes in \textit{Swift}/UVOT filters are reported in Table 2.

\subsection{Radio follow-up observations}
In the radio band, PTF\,10bzf was followed-up by the Allen Telescope Array \citep{Welch2009}, by the Combined Array for Research in Millimeter-wave Astronomy \citep[CARMA,][]{ATEL2473}, and by the Westerbork Synthesis Radio Telescope \citep[WSRT,][]{ATEL2479}. No radio source was detected by any of these telescopes (Table 2).  

The Expanded Very Large Array\footnote{http:$//$www.aoc.nrao.edu$/$evla$/$; The Very Large Array is operated by the National Radio Astronomy Observatory (NRAO), a facility of the National Science Foundation operated under cooperative agreement by Associated Universities, Inc.} \citep[EVLA;][]{Perley2009} provided the deepest upper-limit of $33\,\mu$Jy for a radio counterpart associated with PTF\,10bzf \citep[][and Table 1]{ATEL2483}. 

Starting on 2010 November 27.49, we observed PTF\,10bzf with the EVLA in its C configuration, through an EVLA exploratory program (PI: A. Corsi)\footnote{VLA/10C-227 - ``Late time follow-up of Ic SN PTF\,10bzf''.}. We observed for 2 hrs, at a center frequency of $4.96$ GHz, and with a total bandwidth of 256 MHz. No radio source was detected at the SN position, the corresponding upper-limit is reported in Table 2. A compact source (J1146+5356) near PTF\,10bzf was observed every 6 min for accurate phase calibration, while 3C\,286 was observed at the end of the run for the bandpass and flux density calibration. Data were reduced and imaged using the Astronomical Image Processing System (AIPS) software package.

We also reduced with the AIPS archival data that were taken on 2010 May 21.21 (PI: A. Soderberg), for a 1 hr integration time at 6 GHz (with a 1 GHz bandwidth). We find no source at the position of PTF\,10bzf, and derive a $3\sigma$ upper-limit comparable to the one obtained by \citet{ATEL2483} during the first epoch (see Table 2). 

\section{Comparison with SN\,1998bw and GRB search}
\label{1998bw_comp}

\begin{figure}
\begin{center}
\includegraphics[width=8.5cm]{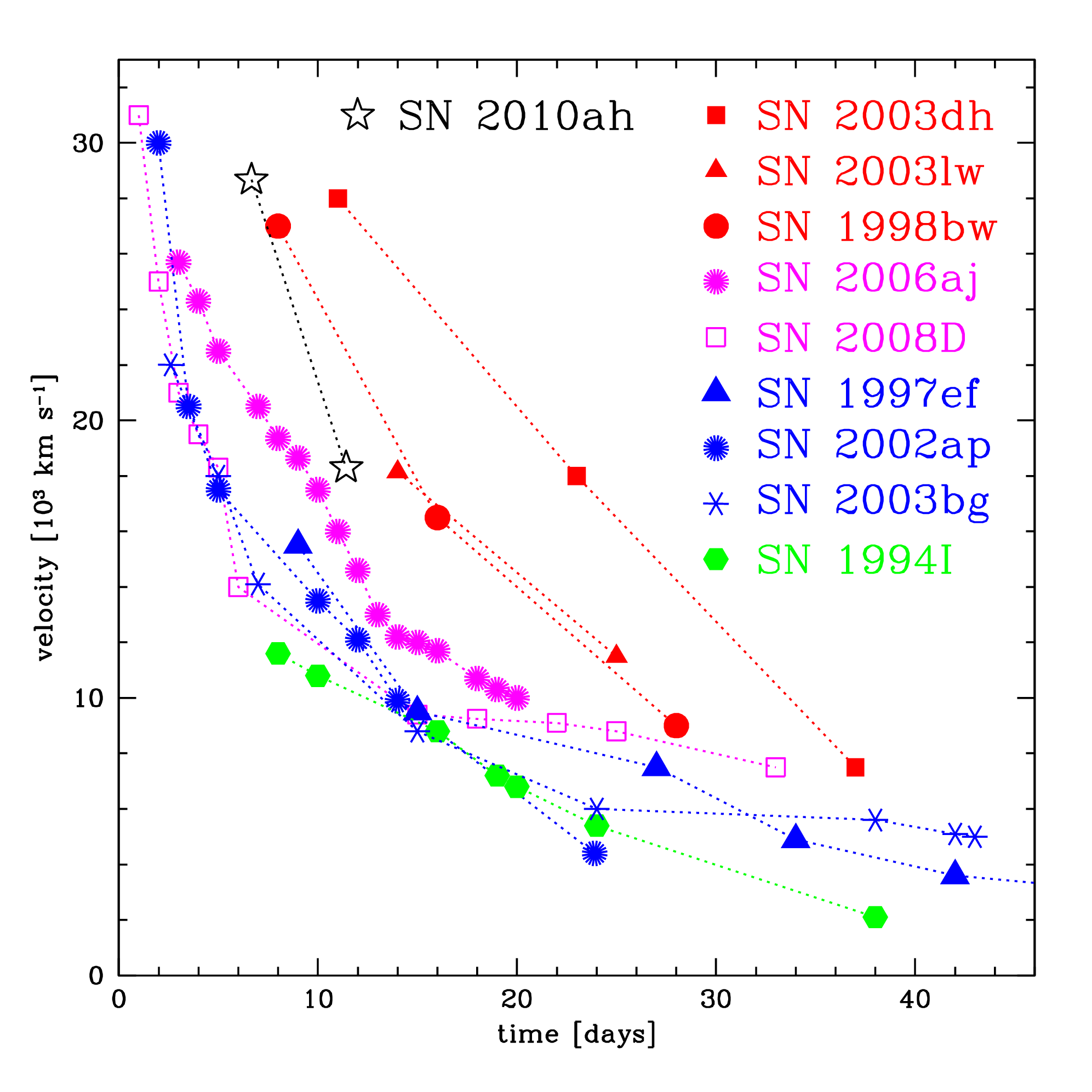}
\caption{The measured \ion{Si}{2} line velocity of PTF\,10bzf (black stars) compared to photospheric velocities {\em derived from spectroscopic modeling} for a number of SNe Ib/c. Red symbols represent GRB-SNe \citep[i.e. SN\,1998bw, SN\,2003dh, SN\,2003lw, see][]{Iwamoto1998,Mazzali2003,Mazzali2006}; magenta is used for XRF/X-ray transients-SNe \citep[i.e. SN\,2006aj and SN\,2008D,][]{Mazzali2006a,Pian2006,Modjaz2009}; blue represent BL-SNe type Ic \citep[SN\,1997ef and SN\,2002ap][]{Mazzali2000,Mazzali2002} or IIb \citep[SN\,2003bg,][]{Mazzali2009}; finally green is used for the ``normal'' SN Ic 1994I \citep{Sauer2006}. 
\label{Cors0121_Fig5}}
\end{center}
\end{figure}

\subsection{Optical emission}
Our photometric monitoring program missed the photometric maximum. However, by performing synthetic photometry on the Gemini spectrum taken on 2010 March 2 (about a week after the discovery), and the Keck spectrum taken on 2010 March 7 (about 12 days after the discovery), we derive $g$-$r$-$i$ magnitudes of PTF\,10bzf at these epochs (see Table 2). As evident from Figure 2, PTF\,10bzf was likely around maximum light during the Gemini observation. The $R$-band apparent magnitude of PTF\,10bzf, as derived from the synthetic photometry on the Gemini spectrum, is $R=18.4\pm0.3$ (AB system, or $R=18.2\pm0.3$ in the Vega
system). This gives us an approximate estimate of the SN absolute magnitude around R-band peak time, of $M_{\rm R}=-18.3\pm0.3$ (AB system, or $M_{\rm R}=-18.5\pm0.3$ in the Vega system).

In Figure 2 and Table 1 we summarize the properties of the spectroscopically confirmed broad-line Ic SNe associated with GRBs; of the broad-line Ic SNe 2003jd and 2002ap, as examples of ``normal'' broad-line Ics; and finally of the broad-line Ic SN\,2009bb \citep{Pignata2010,Soderberg2010}, as prototype of engine-driven SNe with no clear GRB association. 

SNe 1998bw \citep{Iwamoto1998}, 2003dh \citep{Mazzali2003}, and 2003lw \citep{Mazzali2007}, all of which produced GRBs, were brighter and more energetic than other broad-line Ic SNe for which a high-energy signal was either not detected \citep[e.g. SNe 1997ef and SN\,2002ap, see][]{Mazzali2000,Mazzali2002}, or was an XRF \citep[SN 2006aj,][]{Mazzali2006}. As evident from Table 1 and Figure 2, PTF\,10bzf was more luminous than the broad-line Ic SN\,2002ap (with a difference in $R$-band absolute magnitude at peak of $M_{\rm R;10bzf}-M_{\rm R,2002ap}=-1.0\pm0.4$); comparable (within the errors) to the engine driven SN\,2009bb and to the broad-line Ic SN\,2003j around maximum $R$-band light; less luminous than the GRB-associated SN\,1998bw ($M_{\rm R,10bzf}-M_{\rm R,1998bw}=0.9\pm0.3$ at $R$-band maximum light). 

Although the peak epoch is not well known, we can estimate the $^{56}$Ni mass in PTF\,10bzf by interpolating between the $R$-band light-curves of SN\,2002ap \citep[for which $M_{^{56}\rm Ni}\approx0.09$~M$_{\odot}$,][]{Mazzali2002} and SN\,1998bw \citep[for which $M_{^{56}\rm Ni}\approx0.5$~M$_{\odot}$,][]{Nakamura1999,Nakamura2001}. We find that for PTF 10bzf $M_{^{56}\rm Ni}\approx 0.20-0.25$\,M$_{\odot}$. The $^{56}$Ni mass can also be estimated using the empirical relation derived by \citet{Drout2010} from the properties of a large sample of Ic SNe (including broad-line and engine driven SNe):
\begin{equation}
	log(M_{^{56}\rm Ni}/M_{\odot})\simeq-0.41\times M_{\rm R}-8.3\label{peak}
\end{equation}
where $M_{\rm ^{56}Ni}$ is the mass of $^{56}$Ni estimated using the formalism of \citet{Valenti2008} \citep[see also][]{Arnett1982}, and $M_{\rm R}$ is the extinction corrected peak magnitude in the $R$-band \citep[see also][]{Perets2010}. For PTF\,10bzf, using Equation (\ref{peak}), we get $M_{\rm ^{56}Ni}\approx 0.2\,M_{\odot}$, compatible with what obtained by interpolation. This is comparable to SN\,2002aj, associated with XRF\,060218 \citep{Mazzali2006}, and significantly less than all other GRB-associated SNe (see Table 1).

Estimating the kinetic energy requires modeling. However, we can use line velocity as a proxy for it. In Figure \ref{Gemini} and Figure \ref{Keck} we show a spectroscopic sequence of broad-line Ic SNe, from 2002ap to 1998bw. The first (Figure \ref{Gemini}) shows spectra near maximum light, the second (Figure \ref{Keck}) shows spectra obtained about 5~days after maximum. In addition, we show the only near-peak spectrum available of SN 1997ef, a broad-line SN without a GRB \citep{Mazzali2000}. We have ordered the SNe by peak luminosity. This ordering however reveals a sequence also in line velocity and blending. 
The spectra of SN\,1998bw are the most affected by line broadening, as seen for example in the almost complete absence of the re-emission peak near 4500 \AA\ and by the complete blending of the \ion{O}{1} 7773 and the \ion{Ca}{2} IR triplet. PTF\,10bzf is intermediate between SN\,1998bw and the two non-GRB SNe 1997ef and 2002ap in both of these respects. 

We have measured the velocity of the \ion{Si}{2} 6355 \AA\ absorption, which traces reasonably closely the position of the photosphere, in the two available spectra of PTF\,10bzf (see Figure \ref{Gemini} and Figure \ref{Keck}). The results are shown in Figure \ref{Cors0121_Fig5}, where they are compared to the photospheric velocity of a number of Ib/c SNe, with and without a GRB or an XRF. PTF\,10bzf appears to be intermediate in velocity between the GRB-associated SNe (in red) and the non-GRB ones, both broad-lined (blue) and narrow-lined (green). Note that, typically, the measured \ion{Si}{2} line velocity is somewhat larger than the photospheric velocity, so the photospheric velocity of PTF\,10bzf is likely lower than the values shown in Figure 5. This makes PTF\,10bzf more similar to non GRB-associated SNe.

We can use this line velocity measurement, together with an estimate of the width of the light curve of PTF\,10bzf, to derive an approximate estimate of the kinetic energy and ejecta mass for PTF\,10bzf. As evident from Figure 2, the light curve of PTF\,10bzf has to be stretched in time by a factor of $\approx 1.12$ to match the shape of SN\,1998bw light curve.  Moreover, as evident from Figure \ref{Cors0121_Fig5}, the velocity of PTF\,10bzf may be a factor of $\approx 0.9$ smaller than SN 1998bw. Using the formula for the light curve width as a function of ejecta mass and kinetic energy, as delineated by Arnett (1982), we obtain for PTF\,10bzf: $M=(7\pm2)$\,M$_{\odot}$ and $E_{\rm K}=(17\pm5)\times 10^{51}$\,erg.  

PTF\,10bzf thus appears to have an $E_K/M$ ratio of $\approx 2.5$. This is larger than non-GRB SNe like 1997ef (which had $E_K/M\approx 2$) or 2002ap (for which $E_K/M\approx 1.25$), but smaller than SN 1998bw ($E_K/M\approx 3$). We thus conclude that PTF\,10bzf probably lacks the mass and energy required to initiate a GRB. 

\subsection{X-ray upper limit}
 The \textit{Swift}/XRT upper limit for PTF\,10bzf at $8.5$ days, constrains any X-ray source associated with PTF\,10bzf to have a flux $< 1.3\times10^{-14}$~erg~s$^{-1}$cm$^{-2}$, which at the redshift of PTF\,10bzf corresponds to an isotropic luminosity of $< 7.4\times10^{40}$~erg~s$^{-1}$.

The X-ray flux of GRB\,980425 about 1\,day after the burst was $\approx 3 \times 10^{-13}$\,erg\,s$^{-1}$cm$^{-2}$, and it declined at a rate of $\propto t^{-0.2}$ \citep{Nakamura1999,Pian2000}. Rescaling it at $8.5$ days and taking into account SN\,1998bw distance, we obtain an X-ray isotropic luminosity of $\approx 3\times10^{40}$~erg~s$^{-1}$, a factor of $\approx 2.5$ below our X-ray luminosity upper-limit on PTF\,10bzf.

For GRB\,030329, an X-ray flux upper-limit of $< 2.6\times10^{-12}$~erg~s$^{-1}$cm$^{-2}$ was set at an epoch of about 8 days since the burst, using Rossi-XTE observations \citep{Tiengo2003}. From broad-band afterglow modeling, the predicted X-ray flux at 8\,days was $\approx 4\times10^{-13}$~erg~s$^{-1}$cm$^{-2}$ \citep[see Figure 4 in][]{Tiengo2003}, which at the redshift of GRB\,030329 corresponds to an isotropic  luminosity of $\approx 3\times10^{43}$erg~s$^{-1}$, a factor of $\approx 400$ higher than our upper-limit on PTF\,10bzf.

GRB\,031203 had an estimated X-ray flux of $\approx 10^{-13}$~erg~s$^{-1}$cm$^{-2}$ around 8 days \citep[derived extrapolating from the XMM-Newton observations, see Figure 1 in][]{Watson2004}. This corresponds to an isotropic X-ray luminosity of $\approx 3\times10^{42}$erg~s$^{-1}$, a factor of $\approx 40$ higher than our X-ray luminosity upper-limit on PTF\,10bzf.

For GRB/XRF\,060218, \textit{Swift}/XRT measured a flux of $\approx 1.3 \times 10^{-13}$~erg~s$^{-1}$cm$^{-2}$ around 8 days since trigger \citep[see the upper-panel of Figure 2 in][]{Campana2006}. This gives an isotropic X-ray luminosity of $\approx 3\times10^{41}$erg~s$^{-1}$, a factor of $\approx 4$ above PTF\,10bzf upper-limit. 

Finally, \textit{Swift}/XRT observations set an upper-limit of $\approx 3 \times 10^{-14}$~erg~s$^{-1}$cm$^{-2}$ on GRB\,100316D X-ray flux around 8 days since the burst \citep[see Figure 6 in][]{Starling2010}, i.e. an isotropic luminosity upper-limit of $\approx 2\times10^{41}$erg~s$^{-1}$, a factor of $\approx 3$ greater than our upper-limit on PTF\,10bzf.

In conclusion, at $\approx 8$ days since trigger, all GRBs with a spectroscopically confirmed associated SN, had an X-ray isotropic luminosity larger than our upper-limit on PTF\,10bzf, except for GRB\,980425.

\begin{figure}
\begin{center}
\includegraphics[width=8.5cm]{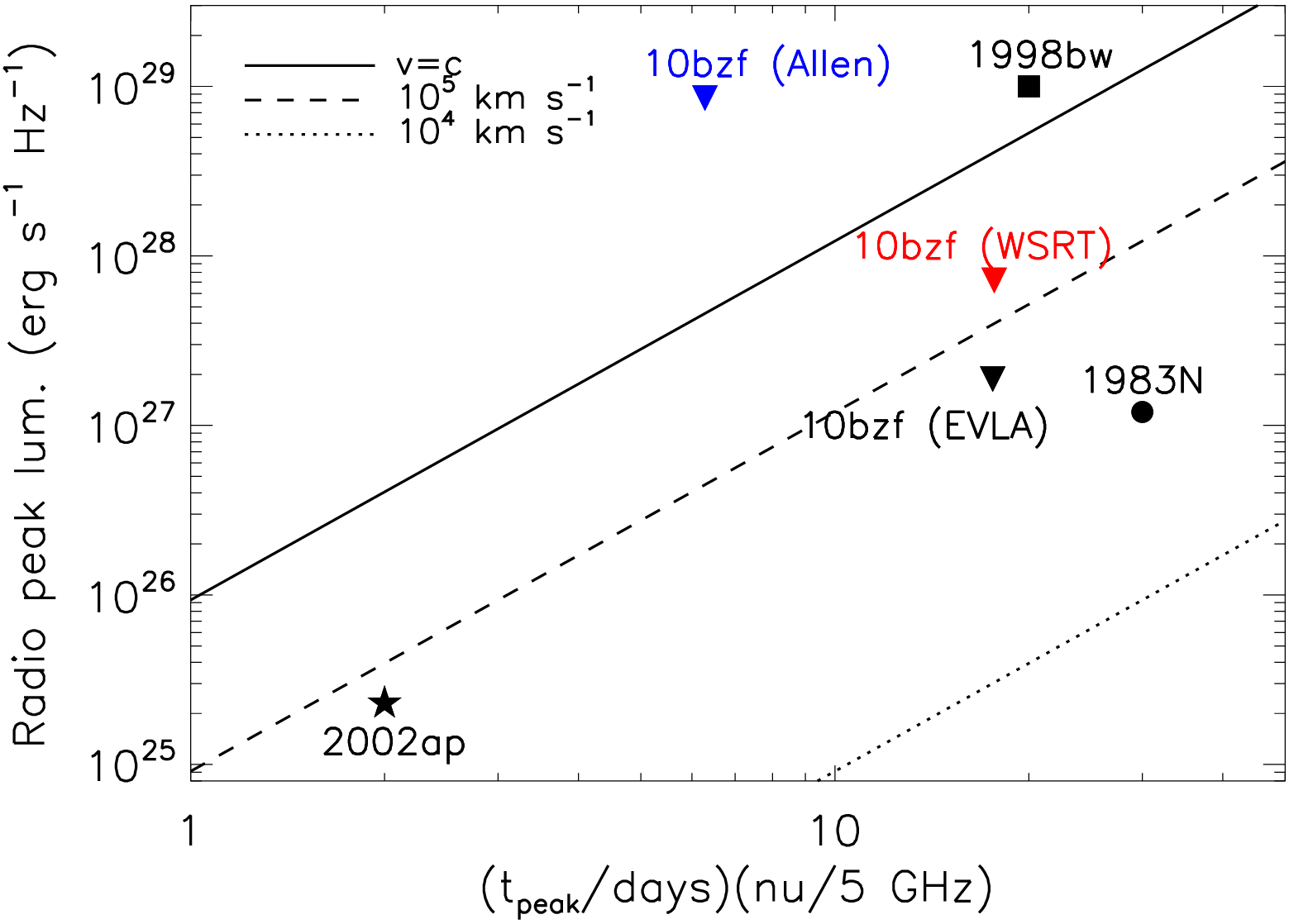}
\caption{Constraints on approximate expansion velocity derived from the early-time Allen telescope (blue triangle), WSRT (red triangle), and EVLA (black triangle) radio upper-limits on PTF\,10bzf (see Table 2 and Equation \ref{eqearly}), compared with similar constraints obtained from the radio detections of SN\,1998bw (black square), SN 1983N (black dot), and SN\,2002ap (black star) \citep[see Figure 2 in][for comparison]{Berger2003}. As evident, SN\,1998bw with v$_r\sim c$ remains an exception. The Allen upper-limit on PTF\,10bzf around its peak time ($\sim 10$ days since discovery, see Table 2) does not exclude a relativistic expansion. However, v$\sim c$ is indeed excluded by the radio upper-limits obtained about a week later by EVLA and WSRT. We stress that while the upper limits do not necessarily measure the peak of the spectrum at the time of the observation, typically time delays of few tens of days relative to the SN explosion reasonably sample the peak. However, if PTF10bzf peaked at few days after the explosion like SN 2002ap, the EVLA and WSRT upper-limits would 
not be deep enough to rule out a relativistic expansion.\label{EARLYradio}}
\end{center}
\end{figure}

\subsection{Early time $(t\lesssim 20$ days) radio upper-limits}
The novelty of SN\,1998bw was in the discovery of prompt radio emission just a few days after GRB\,980425 \citep[][]{Kulkarni1998}. The brightness temperature suggested that the radio photosphere moved relativistically ($\Gamma \geq 2$), with a total energy of $\approx 10^{50}$\,erg \citep{Li1999}, about two orders of magnitude less than the total kinetic energy of the explosion, which was estimated to be of $\approx3\times10^{52}$\,erg \citep{Iwamoto1998}. None of these last features, together with the spatial and temporal association with a GRB, were ever observed for a nearby SN before SN\,1998bw, thus suggesting the presence of a central engine powering this explosion. 

SN\,2003dh, associated with GRB\,030329, showed an isotropic energy of $\approx 4\times10^{52}$\,erg, which leaded to classify this event as an ``hypernova'' \citep{Mazzali2003}. The radio afterglow of GRB\,030329 was $\gtrsim 100$ times more luminous than SN\,1998bw \citep{Berger2003a}. \citet{Berger2003a} modeled such emission as due to the contribution of a second jet component, wider than the one generating the burst itself, and having an isotropic kinetic energy of $\approx 5.6\times10^{51}$\,erg.

SN\,2003lw, associated with GRB\,031203, had a total kinetic energy of $\approx 6\times10^{52}$\,ergs \citep{Mazzali2006}. The spectral peak luminosity in radio was $\approx 10^{29}$\,erg\,s$^{-1}$\,Hz$^{-1}$, about a factor of 100 less luminous than the radio afterglow of GRB\,030329, but comparable to the radio peak luminosity of SN\,1998bw. The kinetic energy in the afterglow component giving rise to the radio emission was estimated to be of $\approx 1.7\times10^{49}$\,erg \citep{Soderberg2004}.

The total kinetic energy of SN\,2006aj was $\approx 2\times10^{51}$\,erg \citep{Mazzali2006a}, smaller than the other GRB-associated SN. The radio afterglow of XRF\,060218 peaked at about 5 days since the trigger, the peak luminosity being about a factor of $10$ less bright than SN\,1998bw at peak \citep{Soderberg2006}. The kinetic energy and Lorentz factor around the time of the radio peak were estimated to be of $\approx 2\times10^{48}$\,erg and $\Gamma\approx2.3$, respectively \citep{Soderberg2006}.

SN\,2010bh, the latest spectroscopically confirmed SN associated with a GRB, had a kinetic energy of $\approx 1.39\times10^{52}$\,erg \citep{Cano2011}. Radio upper-limits at about 2 days since the burst constrain the radio spectral luminosity to be below $8\times10^{27}$\,erg\,s$^{-1}$\,Hz$^{-1}$, a factor of 2-to-10 lower than the radio afterglow luminosities observed for SN\,1998bw, GRB\,031203, and XRF\,060218 on comparable timescales \citep{GCN10533}.

At a redshift of $z=0.0498$, the deepest early time radio upper-limit for PTF\,10bzf was obtained by \citet{ATEL2483}. This sets a limit of on the spectral luminosity of $L_{5\,\rm GHz}<1.6\times10^{27}$\,erg\,s$^{-1}$\,Hz$^{-1}$, at about three weeks after the explosion. This is $\approx 20$ times lower than the radio luminosity observed on a similar timescale for the Ic SN\,1998bw and SN\,2009bb \citep{Soderberg2010}, the only two SNe which showed clear evidence for energetic and mildly-relativistic outflows \citep{ATEL2483}. However, we note that this upper-limit is comparable to the radio luminosity of GRB\,060218 at a similar epoch \citep{Soderberg2006}.

Using the early-time radio upper-limit, we constrain the mean expansion speed v$_r$ of the radio photosphere. Under the reasonable assumption that radio emission arises from a synchrotron spectrum with $\nu_{\rm a}\sim\nu_{\rm p}$ (where $\nu_{\rm a}$ is the self-absorption frequency and $\nu_{\rm p}$ the peak frequency), and under the hypothesis of equipartition, the peak-time $t_{\rm p}$ and peak-luminosity $L_{\rm p}$ of the radio emission directly measure the average expansion speed \citep{Berger2003,Chevalier1998,Kulkarni1998,Scott1977}:
\begin{equation}
	{\rm v_r} \sim 3.1 \times 10^4 \left(\frac{L_{\rm p}}{10^{26} {\rm erg~s}^{-1}}\right)^{17/36}\left(\frac{t_{\rm p}}{10{\rm~days}}\right)^{-1} \left(\frac{\nu_p}{5{\rm\,GHz}}\right)^{-1}{\rm km\,s}^{-1}\,.
\label{eqearly}
\end{equation}

Figure \ref{EARLYradio} shows $L_{\rm 5\,GHz}$ at peak versus $t_{\rm p}\nu_{\rm p}$ for different values of ${\rm v_r}$, compared with the Allen telescope, WSRT and EVLA upper-limits on PTF\,10bzf. Assuming that the radio emission peaked at $\approx 18$\,days, the WSRT and EVLA observations constrain the expansion velocity to be less than $10^{5}$\,km\,s$^{-1}$.

Estimating the expansion velocities for non-detections has the caveat that the radio peak epoch is in reality unknown. However, as noted by \citet{Berger2003}, observations performed at $\sim 10-20$\,days since explosion, sample the radio peak time of Ic SN reasonably well. E.g. SN 1983N peaked in radio at $\sim 30$\,days after explosion, while the radio light-curve of SN\,1998bw showed a double-peak profile, with the first peak around 10 days since explosion, and the second around 30 days after explosion. This justifies our use of radio upper-limits taken within $\approx 3$ weeks since discovery to constrain the expansion speed. We also note, however, that if PTF10bzf peaked at few days after the explosion like SN 2002ap, the EVLA and WSRT upper-limits would not be deep enough to rule out a relativistic expansion.

\subsection{Late time ($t\gtrsim 20$ days) radio upper-limits}
In the context of the fireball model, the emission from an off-axis GRB is expected to be visible in the radio band long past the GRB explosion \citep[at timescales of the order of $\sim 1$ yr, see][]{Levinson2002}. In fact, at sufficiently late times, the relativistic fireball is expected to enter the sub-relativistic phase, during which the jet starts spreading, rapidly intersecting the viewer's line of sight as the ejecta approaches spherical symmetry. 

\begin{figure}
\begin{center}
\includegraphics[width=8.5cm]{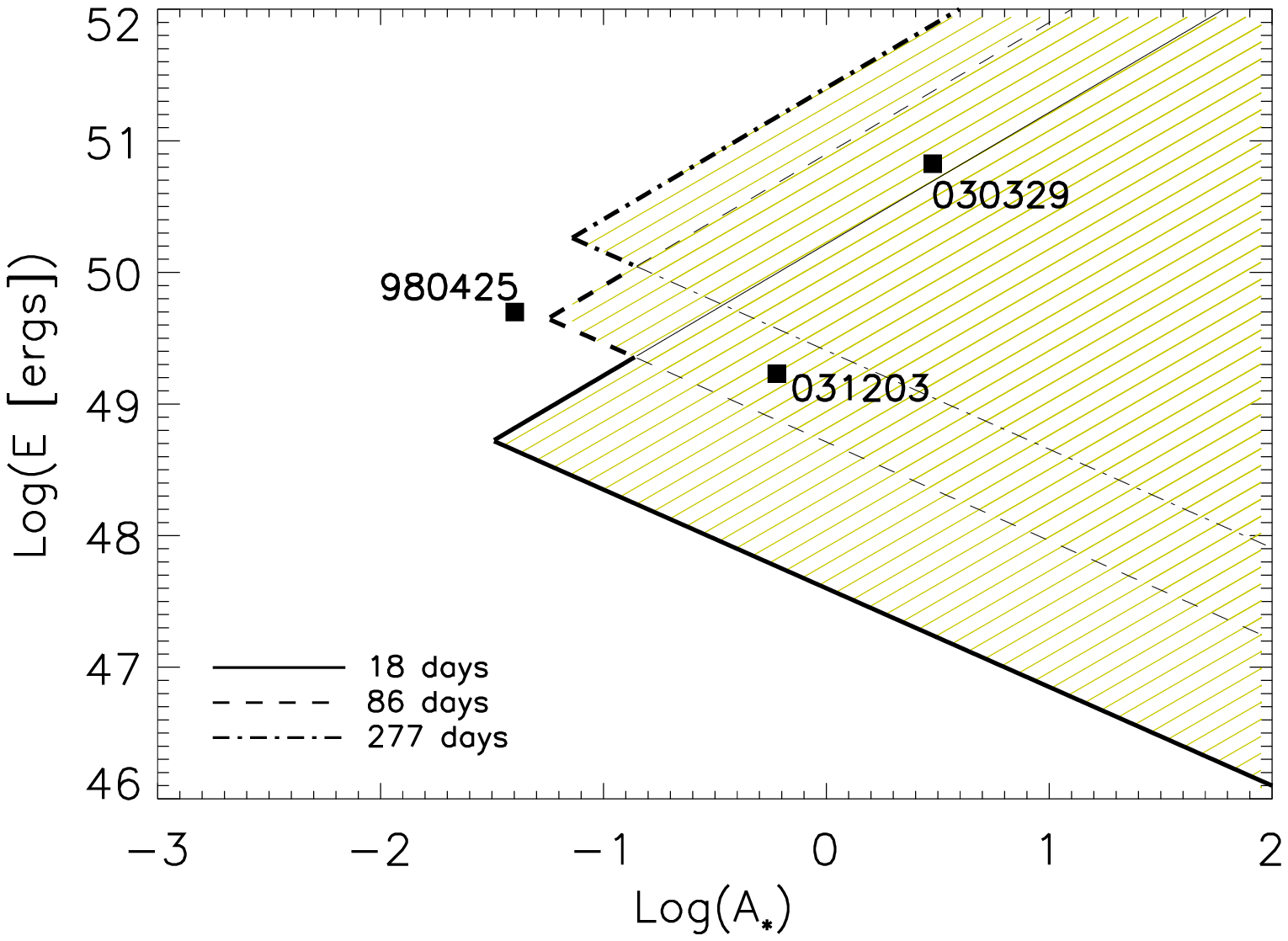}
\caption{Off-axis emission from a GRB explosion associated with PTF\,10bzf. The yellow shadowed regions mark the portion of the energy-density plane excluded by the first, second, and third epoch EVLA observations, respectively (see Table 2).
As evident from the figure, late-time radio observations are fundamental to exclude the higher energy - higher density
region of the energy-density plane. This region contains the most common values of energies and densities
($E_{51} \sim 1-10$ and $A_* \sim 1$) observed for long GRBs. We also show for comparison some of the
parameters estimated from broad-band modeling of GRBs associated with SNe \citep[980425, 030329, and 031203, see][and references therein]{Soderberg2006a,Waxman2004a}\label{LATEradio}.}
\end{center}
\end{figure}

To model the late-time radio emission from an off-axis GRB during the non-relativistic phase, we use the analytical model by \citet{Waxman2004} for a fireball expanding in a wind medium \citep[see e.g.][for the constant ISM case]{Levinson2002}. In this model the radio luminosity is approximated as \citep[see also][]{Perna1998,Livio2000,Pac2001,Granot2002,Berger2003,Soderberg2006a,van2010}:
\begin{eqnarray}
\nonumber	L_{\rm r}\sim 2.1\times10^{29}\left(\frac{\epsilon_{\rm e}\epsilon_{\rm B}}{0.01}\right)^{3/4}\left(\frac{\nu}{10\,{\rm GHz}}\right)^{-\frac{(p-1)}{2}}\left(\frac{t}{t_{\rm NR}}\right)^{-3/2}\\\times~A^{9/4}_{*}
E^{-1/2}_{51}{\rm\,erg\,s}^{-1}\,{\rm Hz}^{-1},
\label{eqlate}
\end{eqnarray}
where $E_{51}$ is the beaming-corrected ejecta energy, $A_*$ defines the circumstellar density in terms of the progenitor mass loss-rate $\dot{M}$ and wind velocity ${\rm v_w}$ such that $\dot{M}/4\pi {\rm v_w}=5\times10^{11}A_*$\,g\,cm$^{-1}$ \citep{Waxman2004,Soderberg2006a}, and:
\begin{equation}
t_{\rm NR} \sim 0.3 \left(\frac{E_{51}}{A_*}\right){\rm\,yr},
\label{tnr}
\end{equation}
is the time of the non-relativistic transition. Note that in the above formulation it is assumed that the fireball spreads sideways at approximately the speed of light. By imposing $t_{\rm NR}\lesssim t_{\rm obs}$ and $L_r\gtrsim L_{\rm obs}$, we can use the EVLA observations of PTF\,10bzf (see Table 2) to exclude the values of (beaming corrected) energy and wind density that would give a radio luminosity above the upper-limits of PTF\,10bzf, at the times of our observations. The exclusion regions obtained in this way are plotted in yellow in Figure \ref{LATEradio}. We can compare these constraints with the energy and density derived from the broad-band afterglow modeling of GRB 980425, GRB 030329 and GRB 031203 \citep[see][and references therein]{Soderberg2006a,Waxman2004a}, also plotted in Figure \ref{LATEradio}. From such a comparison we conclude that a GRB fireball possibly associated with PTF\,10bzf, could not have the same (beaming corrected) energy and wind density of GRB 030329 and GRB 031203. In fact, if this was the case, such a fireball would have been in its non-relativistic (spherical) phase at the times of our radio observations (so that its radio emission would be observable by any off-axis observer), and its radio luminosity would be greater than our upper-limits for PTF\,10bzf. 

We stress that our Figure \ref{LATEradio} only constrains the portion of the 
parameter space where $t_{\rm NR}\lesssim t_{\rm obs}$. This provides an estimate of 
the values of the fireball parameters ruled out by our observations, 
which does not depend critically on the adopted model. 
Constraining the portion of the parameter space where  $t_{\rm NR}\gtrsim t_{\rm obs}$, requires 
detailed modeling. In fact, before $t_{\rm NR}$, the sideways expansion and 
the deceleration of the jet depend on the spatial distribution within 
the jet of the energy density and the Lorentz factor. These distributions 
are poorly constrained by current observations. And, for given 
distributions, an accurate calculation of jet expansion and 
deceleration requires to be carried out numerically.

\subsection{Search for $\gamma$-rays}
\label{GRBsearch}
We searched for a possible GRB in coincidence with PTF\,10bzf using the IPN data. We searched for bursts with a localization error-box including PTF\,10bzf. Since the exact explosion epoch of PTF\,10bzf is not known, we searched in a time-window, extending from 2010 February 12 to 2010 February 23 (the discovery day of PTF\,10bzf). We conservatively extend such a time window to include one full week prior to our last non-detection of PTF\,10bzf. Therefore, our time window starts $\approx 11$\,days before the SN discovery (i.e. $\approx 18$\,days before its $R$-band maximum light). 

For comparison, we note that the engine driven SN\,2009bb reached its maximum light about two weeks after the estimated explosion date \citep{Soderberg2010}, evolving somewhat faster than SN\,1998bw in the pre-peak phase \citep{Pignata2010}. As evident from Figure 2, our last upper-limit on 2010 February 19 ($R>21.3)$ indicates that also PTF\,10bzf probably evolved more rapidly than SN\,1998bw in its pre-peak phase. Using the $R$-band light curve as a proxy for $L(t)$, and assuming a pre-peak luminosity evolution $L(t)\propto (t-t_{0})^{1.8}$ \citep[where $t_0$ is the explosion time, see e.g.][]{Pignata2010}, $t_0$ should be in between $\approx$ 2010 February 17.80 and $\approx$ 2010 February 23.50 (i.e. PTF discovery date) to account for the $\gtrsim 2.44$\,mag drop observed between our last non-detection of PTF\,10bzf, and its discovery. However, we conservatively extend our time interval for the GRB search to 2010 February 12, one week prior to our last non-detection (on 2010 February 19.443).

Between 2010 February 12 and 2010 February 23, a total of 14 confirmed bursts were detected by the spacecrafts of the interplanetary network (IPN\footnote{http:$//$www.ssl.berkeley.edu$/$ipn3$/$}: Mars
Odyssey, Konus-Wind, RHESSI, INTEGRAL (SPI-ACS), \textit{Swift}-BAT, MESSENGER,
Suzaku, AGILE, and \textit{Fermi} (GBM)). Here confirmed means that 
they were observed by more than one detector on one or more
spacecraft, and could be localized at least coarsely. 

The localization accuracies of these 14 bursts varied widely, but none 
of them was found to be consistent with the position of PTF\,10bzf. 
One burst was observed by the \textit{Fermi} GBM alone (error circles with $1\sigma$
statistical-only error radii of 10.5 degrees), while other five were observed by the GBM
and other spacecrafts. Four events were observed by a distant IPN spacecraft, and could be triangulated to
annuli or error boxes with dimensions as small as $\sim 10'$. Five GRBs were observed within the coded field of view of the \textit{Swift} BAT ($3'$
initial localization accuracy), and in some cases by other IPN
spacecrafts as well. 

PTF\,10bzf is not spatially associated with any of these GRBs. The total area of the localizations of the 14 confirmed bursts, containing the 3$\sigma$ error regions, was $\approx 1.3\,sr$. Therefore, there is only $\approx 10$\% chance coincidence between these bursts and PTF\,10bzf.

Next, based on our GRB sample, we can put a limit on the fluence of any GRB associated with PTF\,10bzf. 
We considered three distinct sets of events: IPN bursts, \textit{Fermi} GBM-only bursts, and \textit{Swift}
BAT-only bursts. The IPN is sensitive to bursts with fluences down to
about $6\times10^{-7}$ erg cm$^{-2}$ \citep[50\% efficiency - see][]{Hurley2010}, and observes
the entire sky with a temporal duty cycle close to 100\%.   The \textit{Fermi}
GBM detects bursts down to a $8-1000$~keV fluence of about $4 \times 10^{-8}$
erg cm$^{-2}$, and observes the entire unocculted sky ($\approx 8.8$ sr) with a
temporal duty cycle of more than 80\%.  The weakest burst observed by
the BAT had a 15-150 keV fluence of $6 \times 10^{-9}$ erg cm$^{-2}$, and the BAT
observes a field of view of about 2 sr with a temporal duty cycle of
about 90\%.

If the SN produced a burst below the IPN threshold, and above the \textit{Fermi} one, it is
possible that both \textit{Swift} and \textit{Fermi} did not detect it: considering their
spatial and temporal coverages, the non-detection probabilities are
about $0.86$ and $0.44$, respectively.  Finally, if the SN produced
a burst below the \textit{Fermi} threshold, but above the \textit{Swift} one, the non-detection
probability is about $0.86$.
We note that a burst with an isotropic energy release comparable to that of the sub-energetic GRB\,980425, $E_{\rm iso}\approx 6\times 10^{47}$\,ergs, placed at the distance of PTF\,10bzf, would be observed with a fluence of $\approx 10^{-7}$ erg cm$^{-2}$, which is below the IPN threshold, but above the \textit{Fermi} GBM one.

\section{Conclusion}
\label{conclusion}
We presented the discovery of PTF\,10bzf, a broad-line type Ic SN detected by PTF. We obtained multi-wavelength follow-up observations of this SN, we compared its properties with those of other SNe associated with GRBs, and put limits on any associated GRB using the IPN sample.

While PTF\,10bzf shows some spectral similarities with SN\,1998bw, its $R$-band and radio luminosities are much lower, with no clear evidence for a relativistic expansion speed. The spectral properties of PTF\,10bzf suggest that this SN is intermediate, in terms of explosion kinetic energy, between non-GRB associated SNe like 2002ap or 1997ef, and GRB-associated SNe like 1998bw. A search for $\gamma$-rays using the IPN sample, gives no GRB with a position consistent with this SN, in a time-window extending to a full week prior to our last non-detection of the source. We thus conclude that PTF\,10bzf probably lacks the mass and energy required to initiate a GRB. 

Despite the fact that PTF\,10bzf does not show evidence for being associated with a nearby GRB, the discovery and follow-up of broad-line Ic SNe remains a fundamental tool to investigate the GRB-SN connection. Broad-line Ic SNe are rare, and they are the only type of SNe which have confirmed associations with GRBs. Therefore, it is crucial to study them, and determine how the GRB-associated SNe differ from the other broad-line Ic SN. PTF is a project able to discover and classify about ten broad-line Ic SNe per year, thus allowing to construct a large sample of SNe which are unique broad-line Ic. 

The search for SNe associated with nearby (non $\gamma$-ray triggered) GRBs, is particularly relevant also in the light of multi-messenger astronomy. In an era in which ground-based gravitational wave detectors like LIGO\footnote{www.ligo.org} and Virgo\footnote{www.virgo.infn.it} are approaching their advanced configurations, nearby GRBs represent promising candidates for the detection of gravity waves \citep[e.g.][and references therein]{Meszaros2003,Kokkotas2004,Woosley2006,Piro2007,Corsi2009,Ott2009}. The simultaneous operation of facilities like PTF and LIGO may open, in the forthcoming years, a unique opportunity for this kind of multi-messenger searches \citep[e.g.][and references therein]{Bloom2009,Smith2009,Sta2009,Shawn2010}.\\

\vspace{0.1cm}

\acknowledgments
PTF is a collaboration of Caltech, LCOGT, the Weizmann Institute, LBNL,
Oxford, Columbia, IPAC, and Berkeley.\\
The National Energy Research Scientific Computing Center, which is supported by
the Office of Science of the U.S. Department of Energy under Contract No.
DE-AC02-05CH11231, provided staff, computational resources and data storage for
this project. \\
E. O. O. is supported by an Einstein Fellowship and NASA grants. D. P. is supported
by an Einstein fellowship.\\
The Weizmann Institute PTF partnership is supported in part by grants
from the Israeli Science Foundation (ISF) to A.G. Joint work by the
Weizmann and Caltech groups is supported by a grant from the Binational
Science Foundation (BSF) to A.G. and S.R.K. A.G. acknowledges further support
from an EU/FP7 Marie Curie IRG fellowship and a research grant from the 
Peter and Patricia Gruber Awards. \\
The National Radio Astronomy Observatory is a facility of the National Science Foundation
operated under cooperative agreement by Associated Universities, Inc.\\
The Gemini Observatory is operated by the Association of Universities for Research in Astronomy, Inc., under a cooperative agreement with the NSF on behalf of the Gemini partnership: the National Science Foundation (United States), the Science and Technology Facilities Council (United Kingdom), the National Research Council (Canada), CONICYT (Chile), the Australian Research Council (Australia), Minist\'{e}rio da Ci\^{e}ncia e Tecnologia (Brazil) and Ministerio de Ciencia, Tecnolog\'{i}a e Innovaci\'{o}n Productiva (Argentina).\\
The W.M. Keck Observatory is operated as a scientific partnership among the California Institute of Technology, the University of California and the National Aeronautics and Space Administration. The Observatory was made possible by the generous financial support of the W.M. Keck Foundation.\\
LAIWO, a wide-angle camera operating on the 1-m telescope at the Wise
Observatory, Israel, was built at the Max Planck Institute for Astronomy
(MPIA) in Heidelberg, Germany, with financial support from the MPIA, and
grants from the German Israeli Science Foundation for Research and
Development, and from the Israel Science Foundation.\\ 
LIGO was constructed by the California Institute of Technology and Massachusetts
Institute of Technology with funding from the National Science Foundation.\\
We are grateful to IPN collaborators S. Golenetskii, R. Aptekar, E. 
Mazets, D. Frederiks, and T. Cline for the Konus data;
to M. Briggs and C. Meegan for the \textit{Fermi} data; to T. Takahashi, Y. 
Terada, M. Tashiro, Y. Fukazawa, T. Murakami, M. Ohno,
and K. Makishima for the Suzaku data; to S. Barthelmy, N. Gehrels, H. 
Krimm, and D. Palmer for the Swift data; to D. Golovin,
A. Kozyrev, M. Litvak, A. Sanin, C. Fellows, K. Harshman, and R. Starr 
for the Odyssey data; and to A. von Kienlin and X.
Zhang for the INTEGRAL data.  KH acknowledges support from the following 
NASA sources: NNX09AV61G (Suzaku), NNX10AI23G (Swift),
NNX10AU34G (\textit{Fermi}), and NNX07AR71G (MESSENGER).\\
S.B.C.~acknowledges generous financial assistance from Gary \& Cynthia
Bengier, the Richard \&  Rhoda Goldman Fund, NASA/{\it Swift} grants
NNX10AI21G and GO-7100028, the TABASGO Foundation, and NSF grant
AST-0908886.\\
A.C. and S.R.K. acknowledge partial support from NASA/\textit{Swift} 
Guest Investigator Program Cycle 7 (NNH10ZDA001N).

\bibliography{Cors0121_revisedII}

\begin{thebibliography}{147}
\expandafter\ifx\csname natexlab\endcsname\relax\def\natexlab#1{#1}\fi

\bibitem[{{Abdo} {et~al.}(2009{\natexlab{a}})}]{Abdo2009b}
{Abdo}, A.~A., {et~al.} 2009{\natexlab{a}}, Nature, 462, 331

\bibitem[{{Abdo} {et~al.}(2009{\natexlab{b}})}]{Abdo2009a}
---. 2009{\natexlab{b}}, ApJL, 706, L138

\bibitem[{{Abdo} {et~al.}(2009{\natexlab{c}})}]{Abdo2009c}
---. 2009{\natexlab{c}}, Science, 323, 1688

\bibitem[{{Arcavi} {et~al.}(2010)}]{Arcavi2010}
{Arcavi}, I., {et~al.} 2010, ApJ, 721, 777

\bibitem[{{Arnett}(1982)}]{Arnett1982}
{Arnett}, W.~D. 1982, ApJ, 253, 785

\bibitem[{{Balberg} \& {Loeb}(2011)}]{Balberg2011}
{Balberg}, S., \& {Loeb}, A. 2011, MNRAS, 478, 1715

\bibitem[{{Berger} {et~al.}(2003{\natexlab{a}})}]{Berger2003a}
{Berger}, E., {et~al.} 2003{\natexlab{a}}, Nature, 426, 154

\bibitem[{{Berger} {et~al.}(2003{\natexlab{b}})}]{Berger2003}
---. 2003{\natexlab{b}}, ApJ, 599, 408

\bibitem[{{Bersier} {et~al.}(2004)}]{Bersier2004}
{Bersier}, D., {et~al.} 2004, GRB Coordinates Network, 2544, 1

\bibitem[{{Bloom} {et~al.}(2003){Bloom}, {Frail}, \& {Kulkarni}}]{Bloom2003}
{Bloom}, J.~S., {Frail}, D.~A., \& {Kulkarni}, S.~R. 2003, ApJ, 594, 674

\bibitem[{{Bloom} {et~al.}(2009)}]{Bloom2009}
{Bloom}, J.~S., {et~al.} 2009, ArXiv e-prints 0902.1527

\bibitem[{{Bucciantini} {et~al.}(2008)}]{Bucciantini2008}
{Bucciantini}, N., {et~al.} 2008, MNRAS, 383, L25

\bibitem[{{Bufano} {et~al.}(2010)}]{Bufano2010}
{Bufano}, F., {et~al.} 2010, Central Bureau Electronic Telegrams, 2227, 1

\bibitem[{{Campana} {et~al.}(2006)}]{Campana2006}
{Campana}, S., {et~al.} 2006, Nature, 442, 1008

\bibitem[{{Cano} {et~al.}(2011)}]{Cano2011}
{Cano}, Z., {et~al.} 2011, ArXiv e-prints 1104.5141

\bibitem[{{Carpenter}(2010)}]{ATEL2473}
{Carpenter}, J.~M. 2010, The Astronomer's Telegram, 2473, 1

\bibitem[{{Chevalier}(1976)}]{Chevalier1976}
{Chevalier}, R.~A. 1976, ApJ, 207, 872

\bibitem[{{Chevalier}(1992)}]{Chevalier1992}
---. 1992, ApJ, 394, 599

\bibitem[{{Chevalier}(1998)}]{Chevalier1998}
---. 1998, ApJ, 499, 810

\bibitem[{{Chevalier} \& {Fransson}(2008)}]{Chevalier2008}
{Chevalier}, R.~A., \& {Fransson}, C. 2008, ApJL, 683, L135

\bibitem[{{Chomiuk} \& {Soderberg}(2010)}]{ATEL2483}
{Chomiuk}, L., \& {Soderberg}, A. 2010, The Astronomer's Telegram, 2483, 1

\bibitem[{{Chornock} {et~al.}(2010)}]{Chornock2010}
{Chornock}, R., {et~al.} 2010, ArXiv e-prints, 1004.2262

\bibitem[{{Cobb} {et~al.}(2004)}]{Cobb2004}
{Cobb}, B.~E., {et~al.} 2004, ApJL, 608, L93

\bibitem[{{Corsi} {et~al.}(2010){Corsi}, {Guetta}, \& {Piro}}]{Corsi2010}
{Corsi}, A., {Guetta}, D., \& {Piro}, L. 2010, ApJ, 720, 1008

\bibitem[{{Corsi} \& {M{\'e}sz{\'a}ros}(2009)}]{Corsi2009}
{Corsi}, A., \& {M{\'e}sz{\'a}ros}, P. 2009, ApJ, 702, 1171

\bibitem[{{Della Valle} {et~al.}(2006)}]{DellaValle2006}
{Della Valle}, M., {et~al.} 2006, Nature, 444, 1050

\bibitem[{{Deng} {et~al.}(2005)}]{Deng2005}
{Deng}, J., {et~al.} 2005, ApJ, 624, 898

\bibitem[{{Drout} {et~al.}(2010)}]{Drout2010}
{Drout}, M.~R., {et~al.} 2010, ArXiv 1011.4959

\bibitem[{{Ferrero} {et~al.}(2006)}]{Ferrero2006}
{Ferrero}, P., {et~al.} 2006, A\&A, 457, 857

\bibitem[{{Filippenko}(1997)}]{Filippenko1997}
{Filippenko}, A.~V. 1997, ARA\&A, 35, 309

\bibitem[{{Foley} {et~al.}(2003)}]{Foley2003}
{Foley}, R.~J., {et~al.} 2003, PASP, 115, 1220

\bibitem[{{Frail} {et~al.}(2001)}]{Frail2001}
{Frail}, D.~A., {et~al.} 2001, ApJL, 562, L55

\bibitem[{{Fruchter} {et~al.}(2006)}]{Fruchter2006}
{Fruchter}, A.~S., {et~al.} 2006, Nature, 441, 463

\bibitem[{{Fynbo} {et~al.}(2003)}]{Fynbo2003}
{Fynbo}, J.~P.~U., {et~al.} 2003, \aap, 406, L63

\bibitem[{{Fynbo} {et~al.}(2006)}]{Fynbo2006}
---. 2006, Nature, 444, 1047

\bibitem[{{Gal-Yam} {et~al.}(2008){Gal-Yam}, {Maoz}, {Guhathakurta}, \&
  {Filippenko}}]{GalYam2008}
{Gal-Yam}, A., {Maoz}, D., {Guhathakurta}, P., \& {Filippenko}, A.~V. 2008,
  ApJ, 680, 550

\bibitem[{{Gal-Yam} {et~al.}(2002){Gal-Yam}, {Ofek}, \& {Shemmer}}]{GalYam2002}
{Gal-Yam}, A., {Ofek}, E.~O., \& {Shemmer}, O. 2002, MNRAS, 332, L73

\bibitem[{{Gal-Yam} {et~al.}(2004)}]{GalYam2004}
{Gal-Yam}, A., {et~al.} 2004, ApJL, 609, L59

\bibitem[{{Gal-Yam} {et~al.}(2006{\natexlab{a}})}]{Gal-Yam2006}
---. 2006{\natexlab{a}}, Nature, 444, 1053

\bibitem[{{Gal-Yam} {et~al.}(2006{\natexlab{b}})}]{GalYam2006}
---. 2006{\natexlab{b}}, ApJ, 639, 331

\bibitem[{{Galama} {et~al.}(1998)}]{Galama1998}
{Galama}, T.~J., {et~al.} 1998, Nature, 395, 670

\bibitem[{{Garnavich} {et~al.}(2003)}]{Garnavich2003}
{Garnavich}, P., {et~al.} 2003, IAU Circ., 8114, 1

\bibitem[{{Gezari} {et~al.}(2008)}]{Gerzari2008}
{Gezari}, S., {et~al.} 2008, ApJL, 683, L131

\bibitem[{{Granot} {et~al.}(2002){Granot}, {Panaitescu}, {Kumar}, \&
  {Woosley}}]{Granot2002}
{Granot}, J., {Panaitescu}, A., {Kumar}, P., \& {Woosley}, S.~E. 2002, ApJL,
  570, L61

\bibitem[{{Grassberg} {et~al.}(1971){Grassberg}, {Imshennik}, \&
  {Nadyozhin}}]{Grassberg1971}
{Grassberg}, E.~K., {Imshennik}, V.~S., \& {Nadyozhin}, D.~K. 1971, APSS, 10,
  28

\bibitem[{{Guetta} \& {Della Valle}(2007)}]{Guetta2007}
{Guetta}, D., \& {Della Valle}, M. 2007, ApJL, 657, L73

\bibitem[{{Hjorth} {et~al.}(2003)}]{Hjorth2003}
{Hjorth}, J., {et~al.} 2003, Nature, 423, 847

\bibitem[{{Hook} {et~al.}(2004)}]{Hook2004}
{Hook}, I.~M., {et~al.} 2004, PASP, 116, 425

\bibitem[{{Hurley} {et~al.}(2010)}]{Hurley2010}
{Hurley}, K., {et~al.} 2010, in Italian Phys. Soc. Conf. Proc., ed. {Eds. G.
  Chincarini, P. D'Avanzo, R. Margutti, and R. Salvaterra}, Vol. 102, 529

\bibitem[{{Iwamoto} {et~al.}(1998)}]{Iwamoto1998}
{Iwamoto}, K., {et~al.} 1998, Nature, 395, 672

\bibitem[{{Jordi} {et~al.}(2006){Jordi}, {Grebel}, \& {Ammon}}]{Jordi2006}
{Jordi}, K., {Grebel}, E.~K., \& {Ammon}, K. 2006, A\&A, 460, 339

\bibitem[{{Kamble} {et~al.}(2010)}]{ATEL2479}
{Kamble}, A.~P., {et~al.} 2010, The Astronomer's Telegram, 2479, 1

\bibitem[{{Kasliwal} \& {Cenko}(2010)}]{ATEL2471}
{Kasliwal}, M.~M., \& {Cenko}, S.~B. 2010, The Astronomer's Telegram, 2471, 1

\bibitem[{{Katz} {et~al.}(2010){Katz}, {Budnik}, \& {Waxman}}]{Katz2010}
{Katz}, B., {Budnik}, R., \& {Waxman}, E. 2010, ApJ, 716, 781

\bibitem[{{Katz} {et~al.}(2011){Katz}, {Sapir}, \& {Waxman}}]{Katz2011}
{Katz}, B., {Sapir}, N., \& {Waxman}, E. 2011, ArXiv e-prints, 1103.5276

\bibitem[{{Kawabata} {et~al.}(2003)}]{Kawabata2003}
{Kawabata}, K.~S., {et~al.} 2003, ApJL, 593, L19

\bibitem[{{Kobayashi} \& {M{\'e}sz{\'a}ros}(2003)}]{Meszaros2003}
{Kobayashi}, S., \& {M{\'e}sz{\'a}ros}, P. 2003, ApJ, 589, 861

\bibitem[{{Kokkotas}(2004)}]{Kokkotas2004}
{Kokkotas}, K.~D. 2004, Classical and Quantum Gravity, 21, 501

\bibitem[{{Kulkarni} {et~al.}(1998)}]{Kulkarni1998}
{Kulkarni}, S.~R., {et~al.} 1998, Nature, 395, 663

\bibitem[{{Law} {et~al.}(2009)}]{Law2009}
{Law}, N.~M., {et~al.} 2009, PASP, 121, 1395

\bibitem[{{Levinson} {et~al.}(2002)}]{Levinson2002}
{Levinson}, A., {et~al.} 2002, ApJ, 576, 923

\bibitem[{{Li} \& {Chevalier}(1999)}]{Li1999}
{Li}, Z., \& {Chevalier}, R.~A. 1999, ApJ, 526, 716

\bibitem[{{Liang} {et~al.}(2007){Liang}, {Zhang}, {Virgili}, \&
  {Dai}}]{Liang2007}
{Liang}, E., {Zhang}, B., {Virgili}, F., \& {Dai}, Z.~G. 2007, ApJ, 662, 1111

\bibitem[{{Liang} {et~al.}(2008)}]{Liang2008}
{Liang}, E.-W., {et~al.} 2008, ApJ, 675, 528

\bibitem[{{Liang} {et~al.}(2010)}]{Liang2010}
---. 2010, ApJ, 725, 2209

\bibitem[{{Lithwick} \& {Sari}(2001)}]{Lithwick2001}
{Lithwick}, Y., \& {Sari}, R. 2001, ApJ, 555, 540

\bibitem[{{Livio} \& {Waxman}(2000)}]{Livio2000}
{Livio}, M., \& {Waxman}, E. 2000, ApJ, 538, 187

\bibitem[{{MacFadyen} \& {Woosley}(1999{\natexlab{a}})}]{Mac1999}
{MacFadyen}, A.~I., \& {Woosley}, S.~E. 1999{\natexlab{a}}, ApJ, 524, 262

\bibitem[{{MacFadyen} \& {Woosley}(1999{\natexlab{b}})}]{MacFadyen1999}
---. 1999{\natexlab{b}}, ApJ, 524, 262

\bibitem[{{Malesani} {et~al.}(2004)}]{Malesani2004}
{Malesani}, D., {et~al.} 2004, ApJL, 609, L5

\bibitem[{{Matheson} {et~al.}(2003)}]{Matheson2003}
{Matheson}, T., {et~al.} 2003, ApJ, 599, 394

\bibitem[{{Mazzali} {et~al.}(2009){Mazzali}, {Deng}, {Hamuy}, \&
  {Nomoto}}]{Mazzali2009}
{Mazzali}, P.~A., {Deng}, J., {Hamuy}, M., \& {Nomoto}, K. 2009, ApJ, 703, 1624

\bibitem[{{Mazzali} {et~al.}(2000){Mazzali}, {Iwamoto}, \&
  {Nomoto}}]{Mazzali2000}
{Mazzali}, P.~A., {Iwamoto}, K., \& {Nomoto}, K. 2000, ApJ, 545, 407

\bibitem[{{Mazzali} {et~al.}(2002)}]{Mazzali2002}
{Mazzali}, P.~A., {et~al.} 2002, ApJL, 572, L61

\bibitem[{{Mazzali} {et~al.}(2003)}]{Mazzali2003}
---. 2003, ApJL, 599, L95

\bibitem[{{Mazzali} {et~al.}(2005)}]{Mazzali2005}
---. 2005, Science, 308, 1284

\bibitem[{{Mazzali} {et~al.}(2006{\natexlab{a}})}]{Mazzali2006a}
---. 2006{\natexlab{a}}, Nature, 442, 1018

\bibitem[{{Mazzali} {et~al.}(2006{\natexlab{b}})}]{Mazzali2006}
---. 2006{\natexlab{b}}, ApJ, 645, 1323

\bibitem[{{Mazzali} {et~al.}(2007)}]{Mazzali2007}
---. 2007, ApJ, 670, 592

\bibitem[{{M{\'e}sz{\'a}ros}(2006)}]{Meszaros2006}
{M{\'e}sz{\'a}ros}, P. 2006, Reports on Progress in Physics, 69, 2259

\bibitem[{{Metzger} {et~al.}(2011)}]{Metzger2011}
{Metzger}, B.~D., {et~al.} 2011, MNRAS, 413, 2031

\bibitem[{{Modjaz} {et~al.}(2006)}]{Modjaz2006}
{Modjaz}, M., {et~al.} 2006, \apjl, 645, L21

\bibitem[{{Modjaz} {et~al.}(2008)}]{Modjaz2008}
---. 2008, \aj, 135, 1136

\bibitem[{{Modjaz} {et~al.}(2009)}]{Modjaz2009}
---. 2009, ApJ, 702, 226

\bibitem[{{Nagakura} {et~al.}(2011){Nagakura}, {Ito}, {Kiuchi}, \&
  {Yamada}}]{Nagakura2011}
{Nagakura}, H., {Ito}, H., {Kiuchi}, K., \& {Yamada}, S. 2011, ApJ, 731, 80

\bibitem[{{Nakamura}(1999)}]{Nakamura1999}
{Nakamura}, T. 1999, ApJL, 522, L101

\bibitem[{{Nakamura} {et~al.}(2001){Nakamura}, {Mazzali}, {Nomoto}, \&
  {Iwamoto}}]{Nakamura2001}
{Nakamura}, T., {Mazzali}, P.~A., {Nomoto}, K., \& {Iwamoto}, K. 2001, ApJ,
  550, 991

\bibitem[{{Nakar} \& {Sari}(2010)}]{Nakar2010}
{Nakar}, E., \& {Sari}, R. 2010, ApJ, 725, 904

\bibitem[{{Norris}(2002)}]{Norris2002}
{Norris}, J.~P. 2002, ApJ, 579, 386

\bibitem[{{Ofek} {et~al.}(2007)}]{Ofek2007}
{Ofek}, E.~O., {et~al.} 2007, ApJ, 662, 1129

\bibitem[{{Ofek} {et~al.}(2010{\natexlab{a}})}]{ATEL2470}
---. 2010{\natexlab{a}}, The Astronomer's Telegram, 2470, 1

\bibitem[{{Ofek} {et~al.}(2010{\natexlab{b}})}]{Ofek2010}
---. 2010{\natexlab{b}}, ApJ, 724, 1396

\bibitem[{{Ott}(2009)}]{Ott2009}
{Ott}, C.~D. 2009, Classical and Quantum Gravity, 26, 204015

\bibitem[{{Paczynski}(2001)}]{Pac2001}
{Paczynski}, B. 2001, Acta Astronomica, 51, 1

\bibitem[{{Patat} {et~al.}(2001)}]{Patat2001}
{Patat}, F., {et~al.} 2001, ApJ, 555, 900

\bibitem[{{Perets} {et~al.}(2010)}]{Perets2010}
{Perets}, H.~B., {et~al.} 2010, Nature, 465, 322

\bibitem[{{Perley} {et~al.}(2009)}]{Perley2009}
{Perley}, R., {et~al.} 2009, IEEE Proceedings, 97, 1448

\bibitem[{{Perna} \& {Loeb}(1998)}]{Perna1998}
{Perna}, R., \& {Loeb}, A. 1998, ApJL, 509, L85

\bibitem[{{Pian} {et~al.}(2000)}]{Pian2000}
{Pian}, E., {et~al.} 2000, ApJ, 536, 778

\bibitem[{{Pian} {et~al.}(2006)}]{Pian2006}
---. 2006, Nature, 442, 1011

\bibitem[{{Pignata} {et~al.}(2011)}]{Pignata2010}
{Pignata}, G., {et~al.} 2011, ApJ, 728, 14

\bibitem[{{Piran}(1999)}]{Piran1999}
{Piran}, T. 1999, Phys. Rep, 314, 575

\bibitem[{{Piran}(2004)}]{Piran2004}
---. 2004, Reviews of Modern Physics, 76, 1143

\bibitem[{{Piro} \& {Pfahl}(2007)}]{Piro2007}
{Piro}, A.~L., \& {Pfahl}, E. 2007, ApJ, 658, 1173

\bibitem[{{Podsiadlowski} {et~al.}(2004)}]{Pod2004}
{Podsiadlowski}, P., {et~al.} 2004, \apjl, 607, L17

\bibitem[{{Poznanski} {et~al.}(2002)}]{Poznanski2002}
{Poznanski}, D., {et~al.} 2002, PASP, 114, 833

\bibitem[{{Proga} \& {Begelman}(2003)}]{Proga2003}
{Proga}, D., \& {Begelman}, M.~C. 2003, ApJ, 592, 767

\bibitem[{{Rabinak} \& {Waxman}(2011)}]{Rabinak2011}
{Rabinak}, I., \& {Waxman}, E. 2011, ApJ, 728, 63

\bibitem[{{Racusin} {et~al.}(2009)}]{Racusin2009}
{Racusin}, J.~L., {et~al.} 2009, ApJ, 698, 43

\bibitem[{{Rau} {et~al.}(2009)}]{Rau2009}
{Rau}, A., {et~al.} 2009, PASP, 121, 1334

\bibitem[{{Rhoads}(2003)}]{Rhoads2003}
{Rhoads}, J.~E. 2003, ApJ, 591, 1097

\bibitem[{{Sauer} {et~al.}(2006)}]{Sauer2006}
{Sauer}, D.~N., {et~al.} 2006, MNRAS, 369, 1939

\bibitem[{{Schawinski} {et~al.}(2008)}]{Schawinski2008}
{Schawinski}, K., {et~al.} 2008, Science, 321, 223

\bibitem[{{Schlegel} {et~al.}(1998){Schlegel}, {Finkbeiner}, \&
  {Davis}}]{Schlegel1998}
{Schlegel}, D.~J., {Finkbeiner}, D.~P., \& {Davis}, M. 1998, ApJ, 500, 525

\bibitem[{{Scott} \& {Readhead}(1977)}]{Scott1977}
{Scott}, M.~A., \& {Readhead}, A.~C.~S. 1977, MNRAS, 180, 539

\bibitem[{{Shawhan} {et~al.}(2010){Shawhan}, {LIGO Scientific Collaboration},
  \& {Virgo Collaboration}}]{Shawn2010}
{Shawhan}, P.~S., {LIGO Scientific Collaboration}, \& {Virgo Collaboration}.
  2010, in Bulletin of the American Astronomical Society, Vol.~42, American
  Astronomical Society Meeting Abstracts n. 215, 406.06

\bibitem[{{Smith} {et~al.}(2009)}]{Smith2009}
{Smith}, J., {et~al.} 2009, in Bulletin of the AAS, Vol.~41, Bulletin of the
  American Astronomical Society, 443

\bibitem[{{Soderberg} {et~al.}(2006{\natexlab{a}}){Soderberg}, {Nakar},
  {Berger}, \& {Kulkarni}}]{Soderberg2006a}
{Soderberg}, A.~M., {Nakar}, E., {Berger}, E., \& {Kulkarni}, S.~R.
  2006{\natexlab{a}}, ApJ, 638, 930

\bibitem[{{Soderberg} {et~al.}(2004)}]{Soderberg2004}
{Soderberg}, A.~M., {et~al.} 2004, Nature, 430, 648

\bibitem[{{Soderberg} {et~al.}(2006{\natexlab{b}})}]{Soderberg2006}
---. 2006{\natexlab{b}}, Nature, 442, 1014

\bibitem[{{Soderberg} {et~al.}(2008)}]{Soderberg2008}
---. 2008, Nature, 453, 469

\bibitem[{{Soderberg} {et~al.}(2010)}]{Soderberg2010}
---. 2010, Nature, 463, 513

\bibitem[{{Sollerman} {et~al.}(2006)}]{Sollerman2006}
{Sollerman}, J., {et~al.} 2006, A\&A, 454, 503

\bibitem[{{Stamatikos} {et~al.}(2009)}]{Sta2009}
{Stamatikos}, M., {et~al.} 2009, in astro2010: decadal Survey, ArXiv 0902.3022,
  284

\bibitem[{{Stanek} {et~al.}(2003)}]{Stanek2003}
{Stanek}, K.~Z., {et~al.} 2003, ApJL, 591, L17

\bibitem[{{Starling} {et~al.}(2010)}]{Starling2010}
{Starling}, R.~L.~C., {et~al.} 2010, ArXiv e-prints 1004.2919

\bibitem[{{Thompson} {et~al.}(2004){Thompson}, {Chang}, \&
  {Quataert}}]{Thompson2004}
{Thompson}, T.~A., {Chang}, P., \& {Quataert}, E. 2004, ApJ, 611, 380

\bibitem[{{Tiengo} {et~al.}(2003)}]{Tiengo2003}
{Tiengo}, A., {et~al.} 2003, A\&A, 409, 983

\bibitem[{{Toma} {et~al.}(2007){Toma}, {Ioka}, {Sakamoto}, \&
  {Nakamura}}]{Toma2007}
{Toma}, K., {Ioka}, K., {Sakamoto}, T., \& {Nakamura}, T. 2007, ApJ, 659, 1420

\bibitem[{{Usov}(1992)}]{Usov1992}
{Usov}, V.~V. 1992, Nature, 357, 472

\bibitem[{{Valenti} {et~al.}(2008)}]{Valenti2008}
{Valenti}, S., {et~al.} 2008, MNRAS, 383, 1485

\bibitem[{{van Eerten} {et~al.}(2010){van Eerten}, {Zhang}, \&
  {MacFadyen}}]{van2010}
{van Eerten}, H., {Zhang}, W., \& {MacFadyen}, A. 2010, ApJ, 722, 235

\bibitem[{{Virgili} {et~al.}(2009){Virgili}, {Liang}, \& {Zhang}}]{Virgili2009}
{Virgili}, F.~J., {Liang}, E., \& {Zhang}, B. 2009, MNRAS, 392, 91

\bibitem[{{Watson} {et~al.}(2004)}]{Watson2004}
{Watson}, D., {et~al.} 2004, ApJL, 605, L101

\bibitem[{{Waxman}(2004{\natexlab{a}})}]{Waxman2004a}
{Waxman}, E. 2004{\natexlab{a}}, ApJL, 605, L97

\bibitem[{{Waxman}(2004{\natexlab{b}})}]{Waxman2004}
---. 2004{\natexlab{b}}, Apj, 602, 886

\bibitem[{{Waxman} {et~al.}(2007){Waxman}, {M{\'e}sz{\'a}ros}, \&
  {Campana}}]{Waxman2007}
{Waxman}, E., {M{\'e}sz{\'a}ros}, P., \& {Campana}, S. 2007, ApJ, 667, 351

\bibitem[{{Welch} {et~al.}(2009)}]{Welch2009}
{Welch}, J., {et~al.} 2009, IEEE Proceedings, 97, 1438

\bibitem[{{Wieringa} {et~al.}(2010){Wieringa}, {Soderberg}, \&
  {Edwards}}]{GCN10533}
{Wieringa}, M., {Soderberg}, A., \& {Edwards}, P. 2010, GRB Coordinates
  Network, Circular Service, 10533, 1 (2010), 533, 1

\bibitem[{{Wiersema} {et~al.}(2010)}]{Wiersema2010}
{Wiersema}, K., {et~al.} 2010, GRB Coordinates Network, Circular Service,
  10525, 1 (2010), 525, 1

\bibitem[{{Woosley}(1993)}]{Woosley1993}
{Woosley}, S.~E. 1993, ApJ, 405, 273

\bibitem[{{Woosley} \& {Bloom}(2006)}]{Woosley2006}
{Woosley}, S.~E., \& {Bloom}, J.~S. 2006, ARA\&A, 44, 507

\bibitem[{{Woosley} \& {Heger}(2006)}]{Woosley2006a}
{Woosley}, S.~E., \& {Heger}, A. 2006, ApJ, 637, 914

\bibitem[{{Woosley} \& {Zhang}(2007)}]{Woosley2007}
{Woosley}, S.~E., \& {Zhang}, W. 2007, Royal Society of London Philosophical
  Transactions Series A, 365, 1129

\bibitem[{{Yaron} {et~al.}(2011)}]{Yaron2011}
{Yaron}, S.~E., {et~al.} 2011, in preparation

\bibitem[{{York} {et~al.}(2000)}]{York2000}
{York}, D.~G., {et~al.} 2000, ApJ, 120, 1579

\bibitem[{{Zhang} {et~al.}(2003){Zhang}, {Kobayashi}, \&
  {M{\'e}sz{\'a}ros}}]{Zhang2003}
{Zhang}, B., {Kobayashi}, S., \& {M{\'e}sz{\'a}ros}, P. 2003, ApJ, 595, 950

\end{thebibliography}

\end{document}